\documentclass[aps,prx,reprint,twocolumn,showpacs,floatfix,superscriptaddress]{revtex4-2}

\usepackage{amssymb,amsmath,amstext}                
\usepackage{graphicx}                                               
\usepackage{epstopdf}                                               
\usepackage{color}                                                     
\usepackage{bm}                                                        
\usepackage{appendix}                                              
\usepackage[utf8]{inputenc}

\usepackage[shortlabels]{enumitem}

\usepackage{latexsym}
\usepackage{xcolor}
\usepackage{braket}
\usepackage{ulem}
\definecolor{lblue} {RGB}{51,71,158}
\definecolor{gr4} {RGB}{34,198,34}
\usepackage[colorlinks=true,citecolor=blue,linkcolor=blue,urlcolor=lblue]{hyperref}

\makeatletter\AtBeginDocument{\let\@elt\relax}\makeatother

\begin{document}


\title{Challenges to observation of many-body localization}

\author{Piotr Sierant}
\affiliation{The Abdus Salam International Center for Theoretical Physics, Strada Costiera 11, 34151, Trieste, Italy}
\affiliation{Institute of Theoretical Physics, Jagiellonian University in Krak\'ow,  \L{}ojasiewicza 11, 30-348 Krak\'ow, Poland }
\author{Jakub Zakrzewski}
\email{jakub.zakrzewski@uj.edu.pl}
\affiliation{Institute of Theoretical Physics, Jagiellonian University in Krak\'ow,  \L{}ojasiewicza 11, 30-348 Krak\'ow, Poland }
\affiliation{Mark Kac Complex
Systems Research Center, Jagiellonian University in Krakow, Krak\'ow,
Poland. }

\date{\today}

\begin{abstract}
We study time dynamics of 1D disordered Heisenberg spin-1/2 chain focusing on a regime of large system sizes and a long time evolution. This regime is relevant for observation of many-body localization (MBL), a phenomenon that is expected to freeze the dynamics of the system and prevent it from reaching thermal equilibrium. Performing extensive numerical simulations of the imbalance, a quantity often employed in the experimental studies of MBL, we show that the regime of a slow power-law decay of imbalance persists to disorder strengths exceeding by at least a factor of 2 the current estimates of the critical disorder strength for MBL. Even though we investigate time evolution up to few thousands tunneling times, we observe no signs of the saturation of imbalance that would suggest freezing of system dynamics and provide a smoking gun evidence of MBL. We demonstrate that the situation is qualitatively different when the disorder is replaced by a quasiperiodic potential. In this case, we observe an emergence of a pattern of oscillations of the imbalance that is stable with respect to changes in the system size. This suggests that the dynamics of quasiperiodic systems remain fully local at the longest time scales we reach provided that the quasiperiodic potential is sufficiently strong. Our study identifies challenges in an unequivocal experimental observation of the phenomenon of MBL.
\end{abstract}

\maketitle

\section{Introduction}

Generic isolated quantum many-body systems initialized in an out-of-equilibrium state are expected to approach featureless thermal states
described by the eigenstate thermalization hypothesis \cite{Deutsch91, Srednicki94, Rigol08}. Many-body localization (MBL)  \cite{Gornyi05, Basko06} has been put forward as a mechanism that prevents the approach to equilibrium due to an interplay of interactions and strong disorder. 

The phenomenon of MBL has received a lot of attention over the last decade \cite{Nandkishore15, Alet18, Abanin19}. The MBL phase is characterized by presence of local integrals of motion \cite{Huse14,Ros15,Serbyn13b, Imbrie16, Wahl17, Mierzejewski18, Thomson18} 
that inhibit the transport \cite{Nandkishore15,Znidaric16}, and slow down the spreading of the quantum entanglement \cite{Serbyn13a,iemini2016signatures}. 
MBL has been investigated numerically in disordered spin chains
 \cite{Santos04a, Oganesyan07,Pal10,Luitz15} that map onto spinless fermionic 
 chains, in systems of spinful fermions \cite{Mondaini15, Prelovsek16,Zakrzewski18,Kozarzewski18}  or bosons \cite{Sierant18, Orell19, Hopjan19} and found in systems with  random interactions \cite{Sierant17, Lev16, Li17a} or in various types of quasiperiodic systems \cite{Iyer13, Khemani17, Mace19}.  All those investigations were confirming the belief that MBL is a robust mechanism of ergodicity breaking, that can be expected to occur in a wide class of local, one-dimensional quantum many-body systems provided that a sufficiently strong quenched disorder is present.
 
This belief was challenged in \cite{Suntajs19} where it was argued that MBL might not be stable in the asymptotic sense, i.e. in the limit of an infinite time and system size, and the observations of earlier works indicate only a presence of an MBL regime found at a finite system size and finite times. This lead to an intense debate about the stability of MBL \cite{Sierant20b, Abanin19a, Panda19} and its dynamical properties \cite{Kiefer20, Luitz20, Sels20, Crowley21,Kiefer21}. Despite these works, it is presently unclear whether a stable MBL phase exists much deeper in the MBL regime than it was previously estimated \cite{Morningstar21} or whether there is no stable MBL phase at all \cite{Sels21}. An example of the latter scenario is provided by disordered constrained spin chains which, despite hosting a wide non-ergodic regime at finite system sizes \cite{Chen18pxp} become ergodic in the thermodynamic limit \cite{Sierant21}.

\begin{figure}
  \centering
\includegraphics[width=0.8\linewidth]{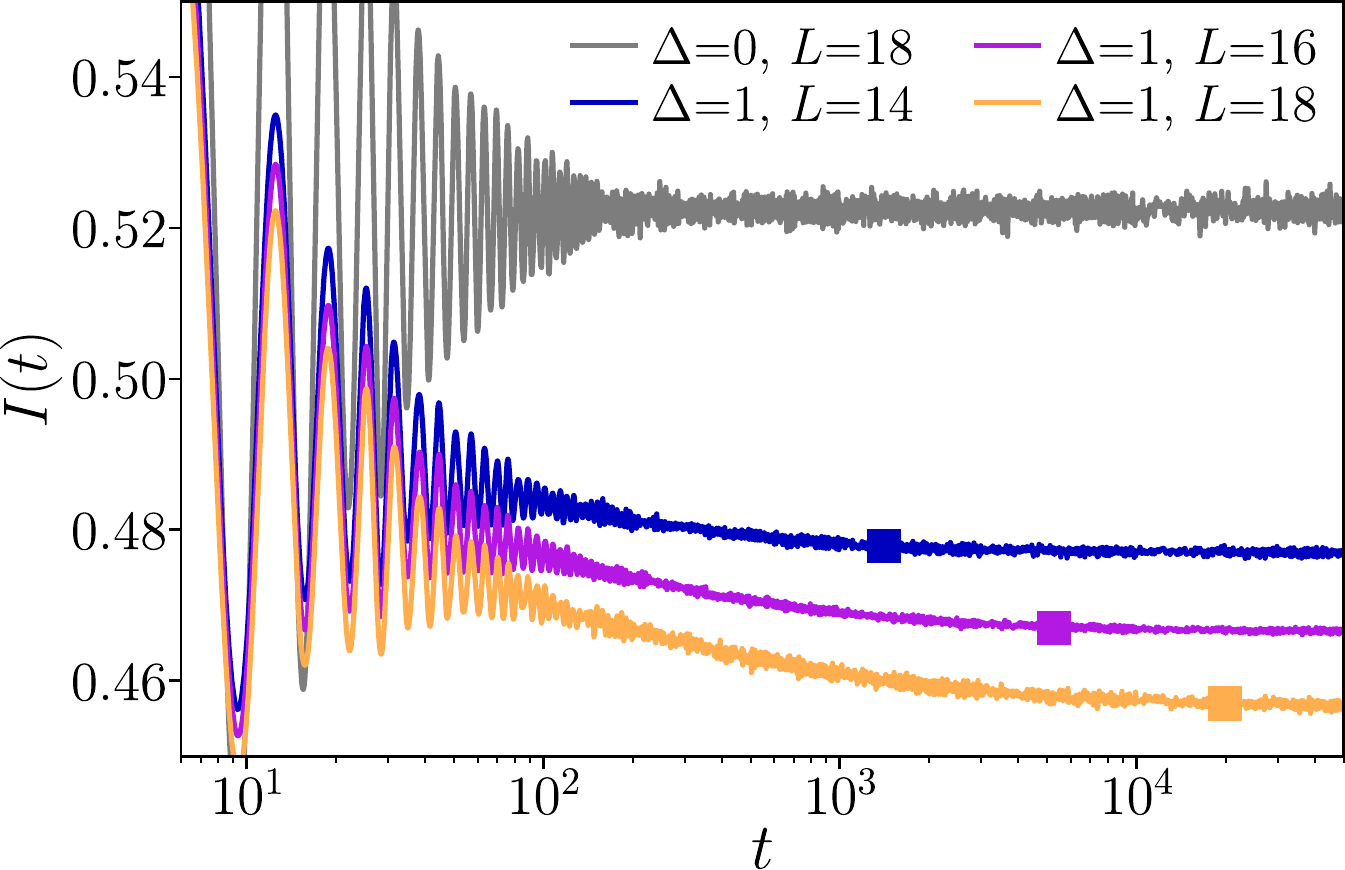}
  \caption{Interactions induce a slow decay of the imbalance $I(t)$  that persists to long times. This is {visualized comparing} results for non-interacting ($\Delta=0$) and interacting ($\Delta=1$) systems. Data for disordered XXZ model \eqref{eq: XXZ} at disorder strength $W=4$. The squares denote the Heisenberg time $t_H$ that scales exponentially with system size $L$.}
  \label{figMoti1}
\end{figure}

The double limit of infinite time and system size is the source of difficulties in establishing the status of MBL. On  one hand, one may investigate properties of eigenstates of many-body systems, that encode the properties of the system at infinite time. However, the eigenstates can be found in an unbiased fashion only for relatively small system sizes $L$ (for instance, for the usually studied spin-1/2 chains, $L \leq 24 $ \cite{Pietracaprina18, Sierant20p}), which does not allow for a fully controlled extrapolation of the results to the thermodynamic limit $L \rightarrow \infty$. On the other hand, tensor network algorithms \cite{Schollwoeck11, Paeckel19} such as {Time Evolving Block Decimation} (TEBD) \cite{Vidal03,Vidal04} or {Time-Dependent Variational Principle} (TDVP) \cite{Haegeman11, Koffel12, Haegeman16, Goto19} allow one to study time evolution of systems comprised of hundreds or even thousands of sites. Unfortunately, the time evolution of many-body systems can be traced faithfully with such algorithms only up to times restricted by the growth of the entanglement in the system. Since, in strongly disordered systems, the entanglement entropy grows only logarithmically in time, maximal times of several hundred tunneling times were achieved in \cite{Doggen18, Zakrzewski18, Chanda19, Chanda20m}. Nevertheless, there is no straightforward way of extrapolating these results to the infinite time limit.

Figure~\ref{figMoti1} illustrates the difficulties in assessing whether the system is ergodic or MBL in a quench experiment. 
It shows the time evolution of the so-called imbalance $I(t)$ for a disordered XXZ spin-1/2 chain (precise definitions are given in the following section). An ergodic system has no memory of its initial state and the imbalance vanishes in the long-time limit: $I(t) \stackrel{t\rightarrow \infty}{\rightarrow} 0$. In contrast, the information about the initial density profile persists indefinitely in the MBL phase in which $I(t) \stackrel{t\rightarrow \infty}{\rightarrow} I_0 >0$. For the non-interacting system ($\Delta=0$) one clearly sees that after initial oscillations, the imbalance saturates to a constant value. Such a behavior allows for a straightforward experimental observation of Anderson localization in the absence of interactions \cite{Billy08, Chabe08}. The main effect of interactions  is that the imbalance decays to much longer times, as exhibited by data for $\Delta=1$. The time scale at which $I(t)$ ceases to decay is of the order of Heisenberg time $t_H$ \cite{Schiulaz19} that is proportional to an inverse of the mean level spacing of the system and hence it is exponentially large in the system size $L$. In a consequence, the data presented in Fig.~\ref{figMoti1} allow us only to conclude that at the considered disorder strength $W=4$, the system is in a {finite time} MBL regime \cite{Morningstar21}. The value of  the imbalance in the $t\rightarrow \infty$ limit is clearly decreasing with {the} system size $L$ and it is impossible to determine from {the} data in Fig.~\ref{figMoti1} whether in the {limit} $L\rightarrow \infty$, $t\rightarrow \infty$ the system remains MBL at $W=4$ or whether the ergodicity is restored.

 The presence of MBL regime has been demonstrated in a number of numerical works as well as in experiments with cold atoms and ions \cite{Schreiber15, Smith16, Luschen17, Kohlert18, Lukin19, Rispoli19, Leonard20}. The aim of this work is to determine whether we can observe unambiguous signatures of the MBL phase in the time evolution of disordered many-body systems. To that end we perform extensive numerical simulations of disordered XXZ spin-1/2 chain and concentrate on the time evolution of density correlation functions.

Let us note that we, on purpose, limit our discussion to short-ranged interactions although MBL has been addressed also for long-range (e.g. dipolar \cite{Yao14,Pino14,Burin15,Burin15b,Deng20}, Ising-type \cite{Hauke15,Singh17,Sous18,Sous19,Botzung19} or cavity-mediated \cite{Sierant19c,Kubala21}) interactions. Similarly we do not address the existence and properties of localization in disorder-free potentials (such as e.g. tilted lattices) - the subject of intensive recent studies \cite{vanNieuwenburg19,Schulz19,Wu19,Taylor20,Guardado20,Khemani20,Doggen20s,Yao20b,Yao21a,Guo20, Chanda20c,Morong21,Scherg21,Yao21}. We want to concentrate on the ``pure'', traditional MBL case. 

The paper is structured as follows. 
In Sec.~\ref{sec:model} we introduce the XXZ spin chain. We provide results for small system sizes and formulate tentative criteria for observation of MBL phase in Sec.~\ref{sec:crit}. Then, we verify whether those criteria are fulfilled by dynamics of the XXZ spin chain in the regime of large disorder strengths and system sizes in Sec.~\ref{sec:imba}. Subsequently, we investigate time evolution of entanglement entropy in that regime in Sec.~\ref{sec:enta}. Finally, instead of random disorder we consider time dynamics of the system with a quasiperiodic potential in Sec.~\ref{sec:imbaQP}. We draw our conclusions in Sec.~\ref{sec:conc}.

\section{Model and observables}
\label{sec:model}

  In this work we concentrate on 1D XXZ spin chain with Hamiltonian given by
\begin{equation}
 H= J \sum_{i=1}^{{L-1}} \left( S^x_{i}S^x_{i+1}+S^y_{i}S^y_{i+1} + \Delta S^z_{i}S^z_{i+1}  \right) + \sum_{i=1}^{L} h_i S^z_i
 \label{eq: XXZ}
\end{equation}
where  $\vec{S}_i$ are spin-1/2 matrices, $J=1$ is fixed as the energy unit,
open boundary conditions are assumed and $h_i \in [-W, W]$ 
are independent, uniformly distributed random variables.
The Jordan-Wigner transformation allows to map XXZ spin chain \eqref{eq: XXZ}, to a 
system of interacting spinless fermions, with the tunneling matrix element equal to $J$ and  
nearest-neighbor interaction strength $\Delta$. This allows to make connection between disordered XXZ model and
optical lattice experiments (as e.g. in \cite{Schreiber15}).
  The random-field XXZ spin chain has been
widely studied in the MBL context, see e.g. \cite{Berkelbach10, Luitz15, Agarwal15, Bera15, Enss17, Bera17, Herviou19, Colmenarez19, Sierant19b, Sierant20, Schiulaz20, TorresHerrera20}. 
Various estimates of the critical disorder strength $W_C$ for the transition to MBL phase include: $W_C\approx 3.7$ \cite{Luitz15}, $W_C\approx 3.8$ \cite{Mace18}, $W_C\approx 4.2$ \cite{Laflorencie20, Chanda19}, $W_C\gtrapprox 5$ \cite{Doggen18, Gray18}, $W_C\approx 5.4$  \cite{Sierant20p}.

Besides the random disorder $h_i \in [-W, W]$, we also consider the case of quasiperiodic (QP) potential, for which $h_j = W^{\mathrm{QP}} \cos(2\pi k j + \phi)$, where $k=(\sqrt{5}-1)/2$ and $\phi$ is a random phase taken from the uniform distribution between  $[0,2\pi]$. The QP potential breaks the translation invariance of the system playing a role similar to disorder and leading to MBL at a critical strong amplitude of the QP potential $W^{\mathrm{QP}}_C$, with various estimates ranging from  $W^{\mathrm{QP}}_C\approx1.5$ \cite{Iyer13, Naldesi16, Setiawan17,BarLev17, Bera17a,Weidinger18} through $W^{\mathrm{QP}}_C \approx 2.4$ \cite{Doggen19, Weiner19}, up to $W^{\mathrm{QP}}_C \approx 4$ \cite{Singh21}. Its important to note that the properties of the transition to MBL phase in QP systems are distinct from the transition in system with random disorder
\cite{Khemani17,Zhang18,Agrawal20, Aramthottil21}.

We analyze dynamics of imbalance 
\begin{equation}
 I(t) = D  \sum_{i=1+l_0}^{L-l_0}   \langle \psi(t) |S^z_i | \psi(t)\rangle \langle \psi | S^z_i |\psi \rangle, 
 \label{eq:imb}
\end{equation}
where $| \psi(t)\rangle = e^{-i H t} |\psi\rangle$, $\ket{\psi}$ is the initial state, the constant $D$ assures that $I(0)=1$, $l_0>0$ diminishes the influence of boundaries (in our calculations we take $l_0=2$). The results are averaged over $n_{\mathrm{real}}$ disorder realizations. As the initial state we take the  N\'eel state with every second spin pointing up and every second spin down $|\psi\rangle = | \uparrow \downarrow \ldots \uparrow \downarrow\rangle$. In the following Section we also take $\ket{\psi}$ as a product state of eigenstates of $S^z_i$ operators with average energy $\bra{\psi} H \ket{\psi}$ being in the middle $10\%$ of the spectrum of $H$ -- we refer to such a choice as to a density correlation function $C(t)$.

We note that other observables, see e.g. \cite{Bera17, Nandy21}, suffer from finite size and finite time limitations similar as \eqref{eq:imb}. Hence, it seems that their behavior is always governed by the broad distributions of relaxation time scales \cite{Vidmar21}, and that is why we concentrate on the very simple observable given by \eqref{eq:imb}, which has another advantage of being directly accessible in experiments with cold atoms \cite{Schreiber15}.

To find the time evolved state $ \ket{\psi(t)}$ we employ Chebyshev expansion of the evolution operator $e^{-i H t}$ \cite{Fehske08}, which allows us  to investigate time evolution of systems of $L\leq 20$ sites up to the Heisenberg time $t_H = 2 \pi / \overline{s} \sim e^{cL}$ (where  $\overline{s}$ is the average level spacing in the middle of the spectrum  {{and $c$ determines the scaling of Hilbert space dimension with system size: for spin-1/s chains $c=\ln 2$}}). For larger system sizes $L=50, 100, 200$ we use a TDVP algorithm, with bond dimension $\chi$, specified later in the text for each $W$ and $L$ considered. In the latter case we focus on relatively large disorder strengths $W\geq 8$ which allows us to investigate time evolution up to a few thousand tunneling times $J^{-1}$.

\section{How to observe an MBL phase?}
\label{sec:crit}

Numerical \cite{Luitz16} as well as experimental \cite{Luschen17} investigations of the imbalance $I(t)$ indicate a presence of a wide regime of disorder strength $W$ in which the imbalance decays according to a power-law $I(t) \sim t^{-\overline \beta}$. 
As a criterion for a transition to MBL, the work \cite{Doggen18} introduced the condition that $\overline  \beta$ vanishing within error bars implies the onset of MBL. The problem with such a criterion is that the error bars on $\overline  \beta$ can be significantly reduced with increasing time of evolution and number of disorder samples, pushing the tentative boundary of MBL to larger and larger disorder strengths. An alternative was put forward in \cite{Chanda20m}, which used a cut-off  $\overline  \beta_{\mathrm {cut} }$ such that $\overline  \beta <\overline  \beta_{\mathrm {cut} }$ implies MBL behavior. The cut-off value of $\overline  \beta_{\mathrm {cut} }$ was taken from a comparison of critical disorder strength estimated from gap ratio statistics as $W_C\approx4$ for system size $L\approx 20$ and the decay rate of imbalance at that system size. 

The latter criterion also runs into problems. If we assume a simplified model of the decay of the imbalance, in which $I(t) \sim t^{-\overline  \beta}$ for $t<t_H$, and then $I(t) = \mathrm{const}$ for $t>t_H$ (which is mildly consistent with data shown in Fig.~\ref{figMoti1}), then the value of the imbalance at infinite time is: $I(\infty)=I(t_H) = e^{-c L \overline  \beta}$. Hence, in order to have a finite value of imbalance in the $t \rightarrow \infty$ the exponent governing decay of imbalance should vanish at least as $ \overline  \beta \sim L^{-1}$. 
Keeping this in mind we now examine the dynamics of the density correlation function $C(t)$ in a system of moderate size $L\leq20$.

   \begin{figure}
   \centering
\includegraphics[width=0.99\linewidth]{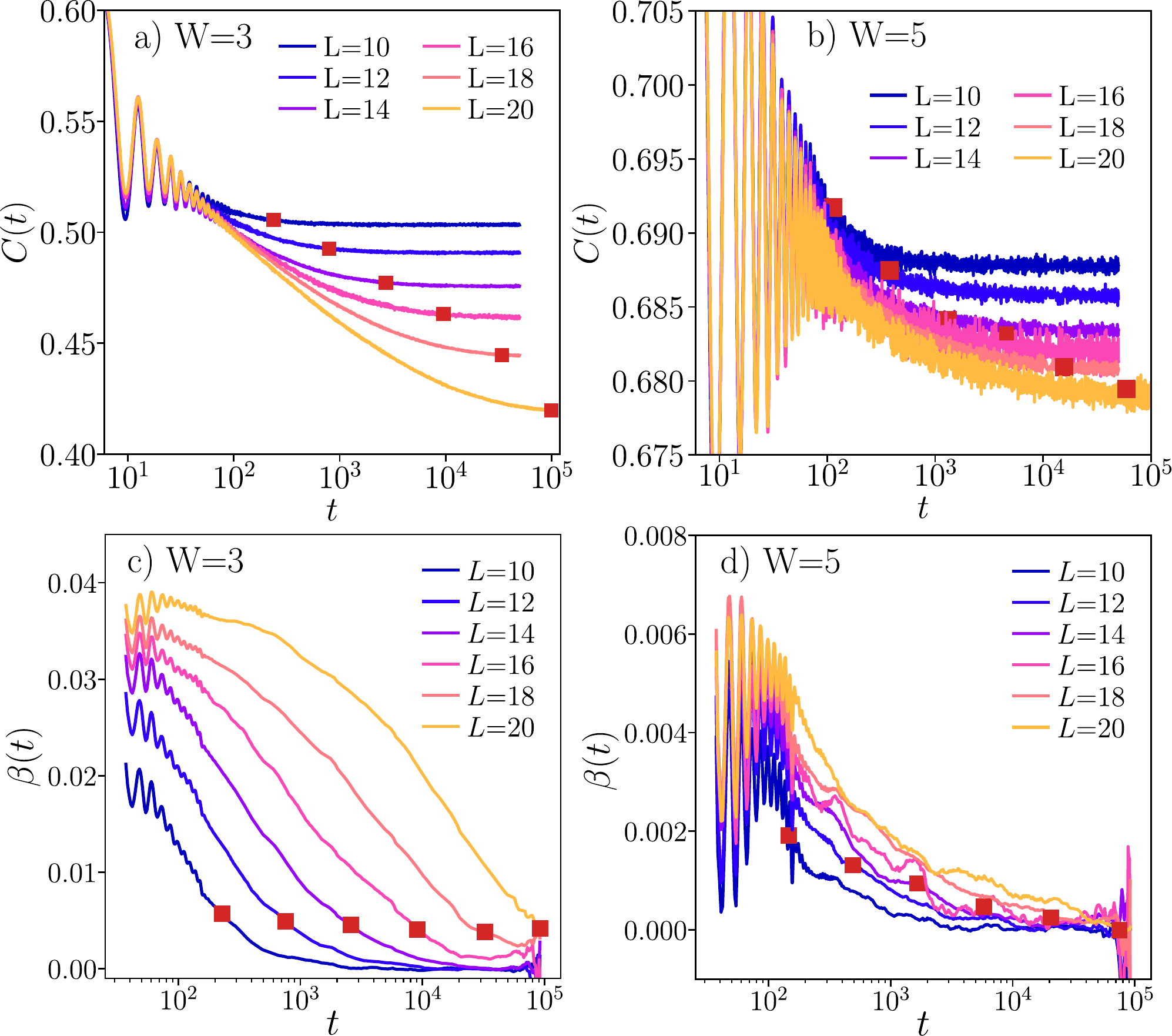}
  \caption{Time evolution of density correlation function $C(t)$ in disordered XXZ model.  Panels a) and b) -- $C(t)$ for various system sizes $L=10,...,L=20$ at disorder strengths $W=3,5$, data averaged over $n_{\mathrm{real}} > 10^4$ disorder realizations. Panels c) and d) -- time evolution of the flowing beta function $\beta(t)$ that locally describes the exponent of decay of $C(t)$.  The red squares denote the Heisenberg time $t_H \sim e^{cL}$.
  }
  \label{figCt}
\end{figure}

Figure~\ref{figCt}a) shows $C(t)$ for disorder $W=3$ for which the XXZ spin chain is in the {ergodic}  phase. 
{The density correlation function, as well as the imbalance are characterized by oscillations at small times due to the coupling between neighboring spins. Those oscillations are gradually damped with time $t$, and the slow decay becomes the main feature of the dynamics of $I(t)$ and $C(t)$. } With an increasing system size, the power-law decay of $C(t)$ persists to longer and longer times, not changing much beyond the Heisenberg time $t_H$. The interaction induced decay of $C(t)$ is evidently getting more abrupt with increasing $L$. The situation is, in fact quite similar for $W=5$ (see Fig.~\ref{figCt}b) ), which, according to the majority of estimates (e.g. \cite{Luitz15,Mace19,Laflorencie20}) is already in the MBL phase. While the decay of $C(t)$ is much slower than for $W=3$, it persist to long-times and the saturation value of $C(t)$ is decreasing with $L$.

To investigate the slow decay of $C(t)$ in more quantitative fashion, we consider a time-dependent $\beta(t)$ function \cite{Doggen19}, that is obtained from the fit $C(t_1)=a t_1^{-\beta(t)}$ in the interval $t\in[t_1, 1.5 t_1]$. The resulting $\beta(t)$ functions are shown in 
Fig.~\ref{figCt} c),d). For $W=3$ we observe that at first, the decay of $C(t)$ is well described by a power-law ($\beta(t)$ is constant) and then the decay gradually slows down, stopping at the time scale approximately order of magnitude larger than $t_H$.  For $W=5$, the slow down of the decay of $C(t)$ occurs at smaller times, however, a non-vanishing $\beta(t)$ up to Heisenberg time signals further, non-negligible decay of the density correlation function.

Results presented in this section show that the correlation functions decay up to Heisenberg time or even longer. Moreover, comparison of results for $W=3$ and $W=5$ indicates that it is hard to propose an accurate phenomenological model for the decay of $C(t)$. Nevertheless, building on intuitions obtained in this section, we conclude that an unambiguous observation of MBL phase should include at least one of the two conditions: 
\begin{enumerate}[(A)]
 \item the value of the exponent $\overline \beta$ that is decreasing with system size as $L^{-1}$ - in such a case even if the power-law decay persists up to the Heisenberg time, the imbalance is non-vanishing in the limit $t\to\infty$; \label{critA}
 \item a decrease of value of $\beta(t)$ {with time $t$} that occurs in a system size independent fashion indicating the saturation of the imbalance {at all experimentally accessible times} beyond a certain time scale. \label{critB}
\end{enumerate}
{The results for small system sizes indicate that if the dynamics of the imbalance satisfies either the criterion \ref{critA} or \ref{critB}, the system is in an asymptotic MBL phase. In that sense, the conditions \ref{critA} and \ref{critB} can be thought of as conditions \textit{sufficient} for the observation of MBL phase. The conditions \ref{critA} and \ref{critB} must be verified with care and their fulfillment is not in a \textit{strict} sense a proof for a stable MBL phase: a system that satisfies either of them could still be ergodic. For instance, one may imagine a decrease of $\beta(t)$ in time in a system size independent fashion below a certain (large from the experimental perspective) time scale combined with an onset of a fast decay of imbalance beyond a certain larger time scale. Nevertheless, such scenarios seem to be ruled out by the results for small system sizes and for that reason we treat  the conditions \ref{critA} and \ref{critB} as sufficient for observation of MBL phase. 
At the same time,  we would like to note that neither of the conditions is a \textit{necessary} criterion for an observation of MBL phase. Other scenarios in which the system breaks ergodicity can be envisioned. For instance, the imbalance may behave in a non-monotonous in time manner with a non-zero infinite time average in the large system size limit, disallowing the analysis of $I(t)$ with a power-law decay.   
}

{With those remarks in mind}, we now turn to an analysis of time dynamics of large systems in the strong-disorder, long-time regime, which seems to be the  most suitable one to find  signatures of the \textit{MBL phase}. {The criteria \ref{critA} and \ref{critB} will be the guiding principles of our analysis.} {First, however, let us briefly consider a non-interacting system.}

 \begin{figure}
\vspace{1cm}
  \centering
\includegraphics[width=0.99\linewidth]{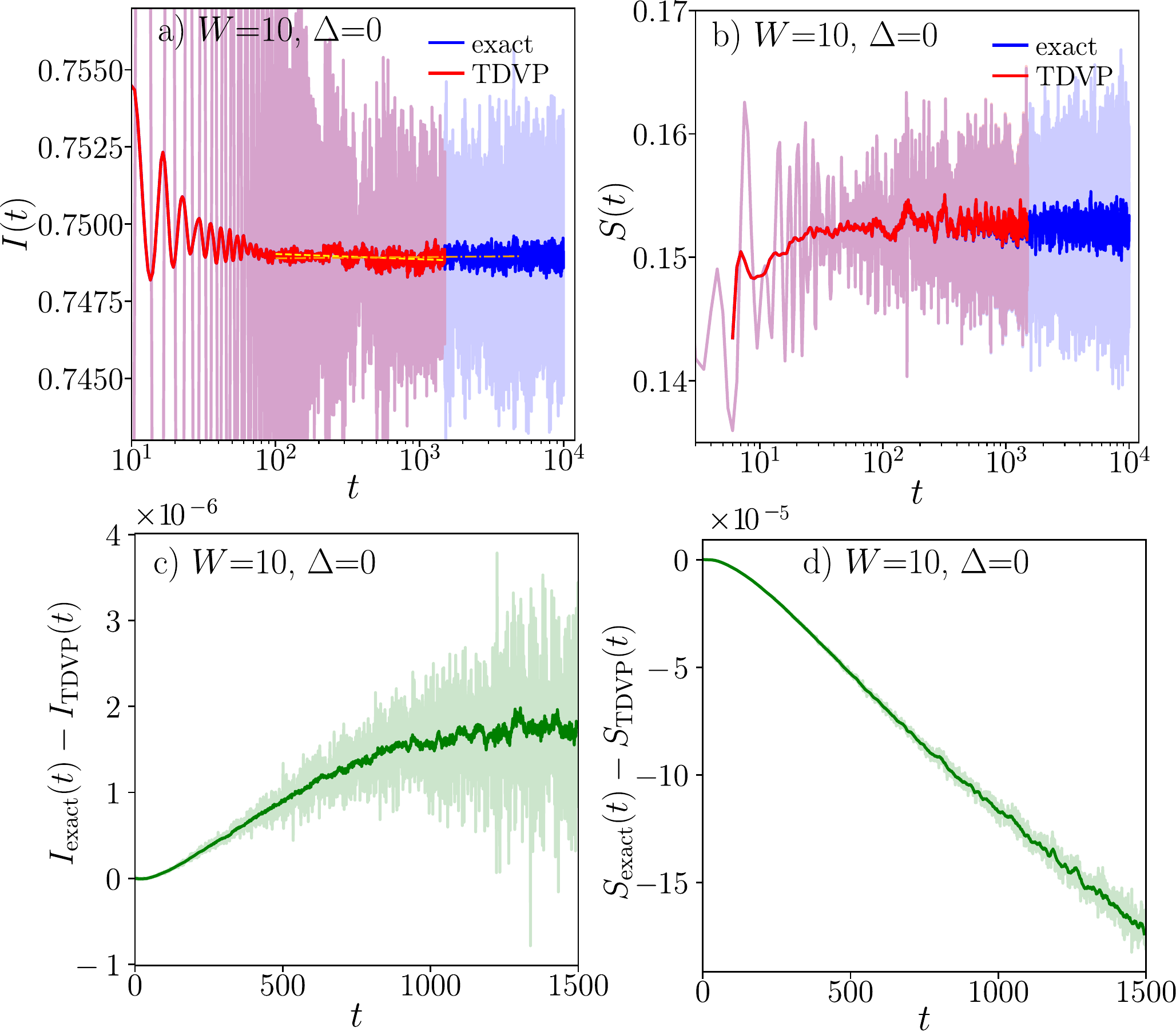}
  \caption{Comparison of the
  time evolution  for noninteracting system between exact propagation and TDVP approximate algorithm ($L/2$ fermions for the system size $L=50$ at disorder strength $W=10$). Top: the imbalance, $I(t)$ (left) and the entanglement entropy in the middle of the chains, $S(t)$ (right) obtained in exact propagation (blue curves extending to larger times) and via TDVP (lighter, orange line). Bottom shows the difference between exact and TDVP results for imbalance (left) and entropy (right).}
  \label{figAnd}
\end{figure}

  \begin{table}
\caption{\label{tab:table0}%
Details of numerical simulations for $W=10$ and $\Delta=0$: system size $L=50$, maximal time reached in time evolution $t_{\mathrm{max} }$, the bond dimension $\chi$ (not displayed for the exact numerical calculation), number of disorder realizations $n_{\mathrm{ real } }$, and the exponent $ \overline \beta $ obtained from the fit $I(t) \sim t^{- \overline \beta} $ in interval $t\in[100,t_{\mathrm{max} }]$. {The error of $ \overline \beta$ is estimated by resampling over the disorder realizations (here, as well in the rest of this manuscript). }
}
\begin{ruledtabular}
\begin{tabular}{lcccc}
{ } &
\textrm{$t_{\mathrm{max} }$}&
\textrm{$\chi$}&
\multicolumn{1}{c}{$n_{\mathrm{ real } }$ }&
\textrm{$\overline \beta$}\\
\colrule 
$L$=50  & 1500 & 128 & 1000 & $(0.97 \pm 1.12)\cdot 10^{-4} $\\
$L$=50  & 1500 & - & 1000 & $(0.96 \pm 1.12)\cdot 10^{-4} $\\
$L$=50  & 5000 & - & 1000 & $(-0.23 \pm 0.43)\cdot 10^{-4} $\\
\end{tabular}
\end{ruledtabular}
\end{table}

 \section{Non-interacting test case}
 \label{sec:nonint}
 
 We consider now the Hamiltonian \eqref{eq: XXZ} and set $\Delta=0$  which via Jordan-Wigner transformation maps to a set of non-interacting spinless fermions in a random on-site potential $h_i$. This model is known to be Anderson localized \cite{Anderson58} for an arbitrary amplitude of the disorder $W$. Since the model is non-interacting, we calculate the time evolution of an initial state $\ket{\psi}$ in numerically exact fashion in time polynomial in system size {(see Appendix  \ref{app:free})}.
 As the initial state we take the N\'eel state $|\psi\rangle = | \uparrow \downarrow \ldots \uparrow \downarrow\rangle$. 
The obtained time evolved imbalance $I(t)$ provides a reference for our approximate
 time propagation using TDVP.  
 Note that while in the non-interacting case obtaining the exact solutions for arbitrary disorder realization is a straightforward task, this is not so for TDVP -- in the latter case the
 algorithm keeps track of a matrix product state that belongs to the full many-body Hilbert space in a manner similar to the interacting case.

The TDVP algorithm used is described in detail in the Appendix~\ref{app:num}. 
{The convergence of TDVP crucially relies on a value of the bond dimension $\chi$. The time evolved state $\ket{\psi_{\chi}(t)}$ obtained with TDVP becomes a better and better approximation of the exact time evolved state $\ket{\psi(t)}$ as $\chi$ increases. However, the simulation cost increases with the value of the bond dimension as $\chi^3$. Hence, one has to choose the value of $\chi$ such that the observables of interest are converged with the bond dimension, i.e. do not change with increase of $\chi$ so that one can safely assume that their value approximates well the value in the exact time evolved state $\ket{\psi(t)}$. For the interacting model \eqref{eq: XXZ} we present details on the convergence of results with the bond dimension $\chi$ in Appendix~\ref{app:num}. In the remainder of this section we compare the exact solution $\ket{\psi(t)}$ for the non-interacting case ($\Delta=0$) with the time evolved state obtained with TDVP.
}

{
For our test we take disorder amplitude $W=10$ and propagate the N\'eel state up to time $t_{\mathrm{max}} =1500$ for 1000 disorder realizations using TDVP with bond dimension $\chi=128$. Fig.~\ref{figAnd}a) compares the obtained imbalance $I(t)$ with the result of exact numerical solution for non-interacting model. The exact solution and TDVP result agree very well up to $t_{\mathrm{max}}=1500$ reach in TDVP simulation. The exact imbalance typically exceeds the TDVP result, the difference, shown in Fig.~\ref{figAnd}c) , grows in time and saturates around $t=800$ at $2\cdot10^{-6}$. The TDVP slightly underestimates the imbalance in agreement with the findings of \cite{Chanda19}. Nevertheless, both TDVP as well as the exact results show that the exponent $\overline \beta$ governing the decay of the imbalance is vanishing within the estimated error bars as shown in Tab.~\ref{tab:table0}. The vanishing $\overline \beta$  fulfills trivially the criterion \ref{critA} for observation of localization.
}

We also calculated entanglement entropy $S(t)$ for a bipartition of the lattice into subsystems $A$ and $B$ of length $L/2$:
\begin{equation}   
 S(t) = -  \mathrm{Tr}_A[\rho_A \ln  \rho_A],
\end{equation}
where $\rho_A = \mathrm{Tr}_B  \ket{\psi(t)}\bra{\psi(t)} $, $\mathrm{Tr}_{C}$ denotes trace with respect to degrees of freedom of subsystem $C$ and $\ket{\psi(t)}$ is the state of the system.
The entanglement entropy $S(t)$ is shown in Fig.~\ref{figAnd}~b). We observe that after an initial increase, the entropy oscillates around a constant value - similarly to the imbalance $I(t)$.  As Fig.~\ref{figAnd}~d)shows, TDVP slightly overestimates the entanglement entropy. The ratio between the error of TDVP simulation and the value of the observable is roughly two orders of magnitude larger than for the imbalance. Nevertheless, the results from TDVP and the numerically exact simulation practically overlap showing that TDVP provides a reliable information about the entanglement entropy growth.

Encouraged by this comparison we shift towards the interacting case for which a comparison with the exact dynamics is not possible. There, we  necessarily rely on self-consistency tests of our simulations described in Appendix \ref{app:num}.

\section{Time evolution of imbalance at strong disorder in large systems}
\label{sec:imba}

\begin{figure}
\vspace{1cm}
  \centering
\includegraphics[width=0.99\linewidth]{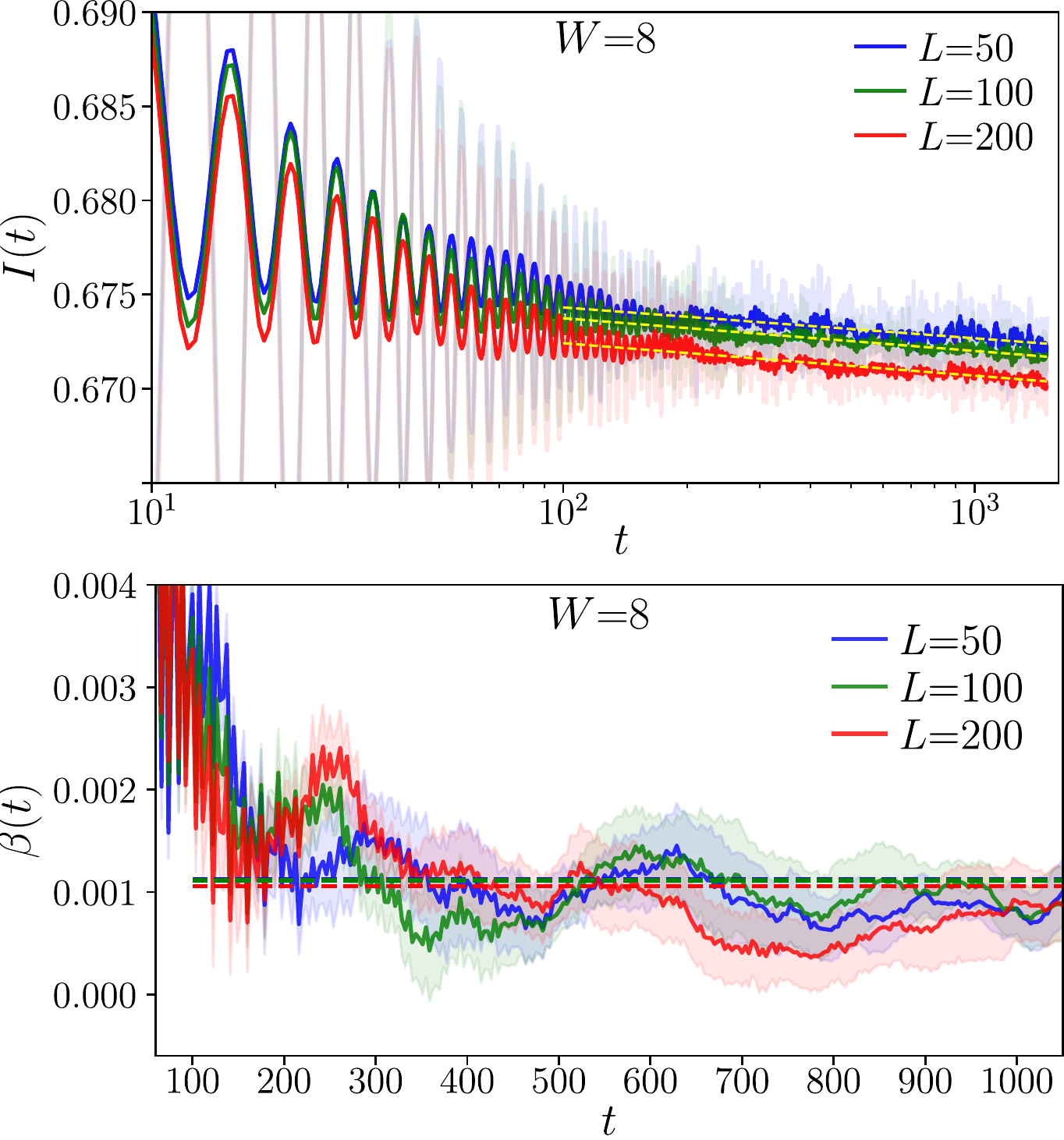}
  \caption{
  Time evolution of imbalance $I(t)$ for systems of size $L=50,100,200$ at disorder strength $W=8$, details of the simulations and fits are given in Tab.~\ref{tab:table1}. Top: the shaded lines denote $I(t)$ whereas the solid lines denote a running overage of $I(t)$ over window $(t-25, t+25)$, dashed lines denote power-law fits $I(t) \sim t^{- \overline \beta} $ in time interval $t\in[100,1500]$. Bottom: the running beta function $\beta(t)$, dashed lines show the value of $\overline \beta$, {the error of $ \beta(t)$ is estimated by resampling over the disorder realizations (here, as well in the rest of this manuscript). }
 }
  \label{figRDW8}
\end{figure}

Taking into account the various estimates of the critical disorder strength $W_C$ for transition to MBL phase, discussed in Sec.~\ref{sec:model}, we fix the disorder {amplitude at} $W=8$ and $W=10$. Such disorder strengths, according to the aforementioned estimates of $W_C$, are expected to {lay} significantly above the transition to the MBL phase.

  \begin{table}
\caption{\label{tab:table1}%
Details of numerical simulations for $W=8$: system size $L$, maximal time reached in time evolution $t_{\mathrm{max} }$, the bond dimension $\chi$, number of disorder realizations $n_{\mathrm{ real } }$, and the exponent $ \overline \beta $ obtained from the fit $I(t) \sim t^{- \overline \beta} $ in interval $t\in[100,1500]$. {The error of $ \overline \beta$ is estimated by resampling over the disorder realizations (here, as well in the rest of this manuscript). }
}
\begin{ruledtabular}
\begin{tabular}{lcccc}
{ } &
\textrm{$t_{\mathrm{max} }$}&
\textrm{$\chi$}&
\multicolumn{1}{c}{$n_{\mathrm{ real } }$ }&
\textrm{$\overline \beta$}\\
\colrule 
$L$=50  & 1500 & 128 & 4000 & $(10.03 \pm 1.23)\cdot 10^{-4} $\\
$L$=100  & 1500 & 128 & 2000 &$(11.07 \pm 0.97)\cdot 10^{-4} $\\
$L$=200  & 1500 & 160 & 1000 &$(11.03 \pm 0.81)\cdot 10^{-4}$
\end{tabular}
\end{ruledtabular}
\end{table}

The evolution of imbalance $I(t)$ for $W=8$ is shown in Fig.~\ref{figRDW8} whereas the details of numerical simulations are shown in Tab.~\ref{tab:table1}. After an initial transient decay and oscillations that last up to $t\approx 100$, we observe a slow but steady monotonic decrease of $I(t)$ that persists up to the largest time $t_{\mathrm{max}}=1500$ reached in the simulation. The value of $t_{\mathrm{max}}$ is not sufficiently large to unambiguously pin-point the functional form of the decay of $I(t)$. Nevertheless, we observe that the imbalance is well fitted by a power-law decay $I(t) \sim t^{- \overline \beta} $ in the interval $t\in[100,1500]$. The values of the exponent $\overline \beta$, shown in Tab.~\ref{tab:table1}, are positive confirming that the slow decay of $I(t)$ is present {(for a discussion of the stability of the value of $\overline \beta$ with respect to the choice of the fitting interval see Appendix.~\ref{app:tmin})}. Moreover, within the estimated error bars, the values of $\overline \beta$ are the same for system sizes $L=50, 100, 200$, indicating clearly that the condition \ref{critA} for the observation of MBL phase is not met at $W=8$. 

To check whether the condition \ref{critB} is fulfilled, we consider the flowing beta function $\beta(t)$ obtained from fitting $I(t_1)=a t_1^{-\beta(t)}$ in the interval $t\in[t_1, 1.5 t_1]$. The result, shown in the bottom panel of Fig.~\ref{figRDW8}, indicates that the decay of the imbalance slows down considerably for $t{\approx} 150$. However, beyond that time the value of the  $\beta(t)$ oscillates around the exponent $\overline \beta$. Therefore, we see no traces of slowing-down of the decay of imbalance at $W=8$.

In conclusion, for $W=8$, neither the criterion \ref{critA} nor \ref{critB} is fulfilled. Hence, we proceed to repeat our analysis for larger disorder strength $W=10$.
  \begin{figure}
  \centering
\includegraphics[width=0.99\linewidth]{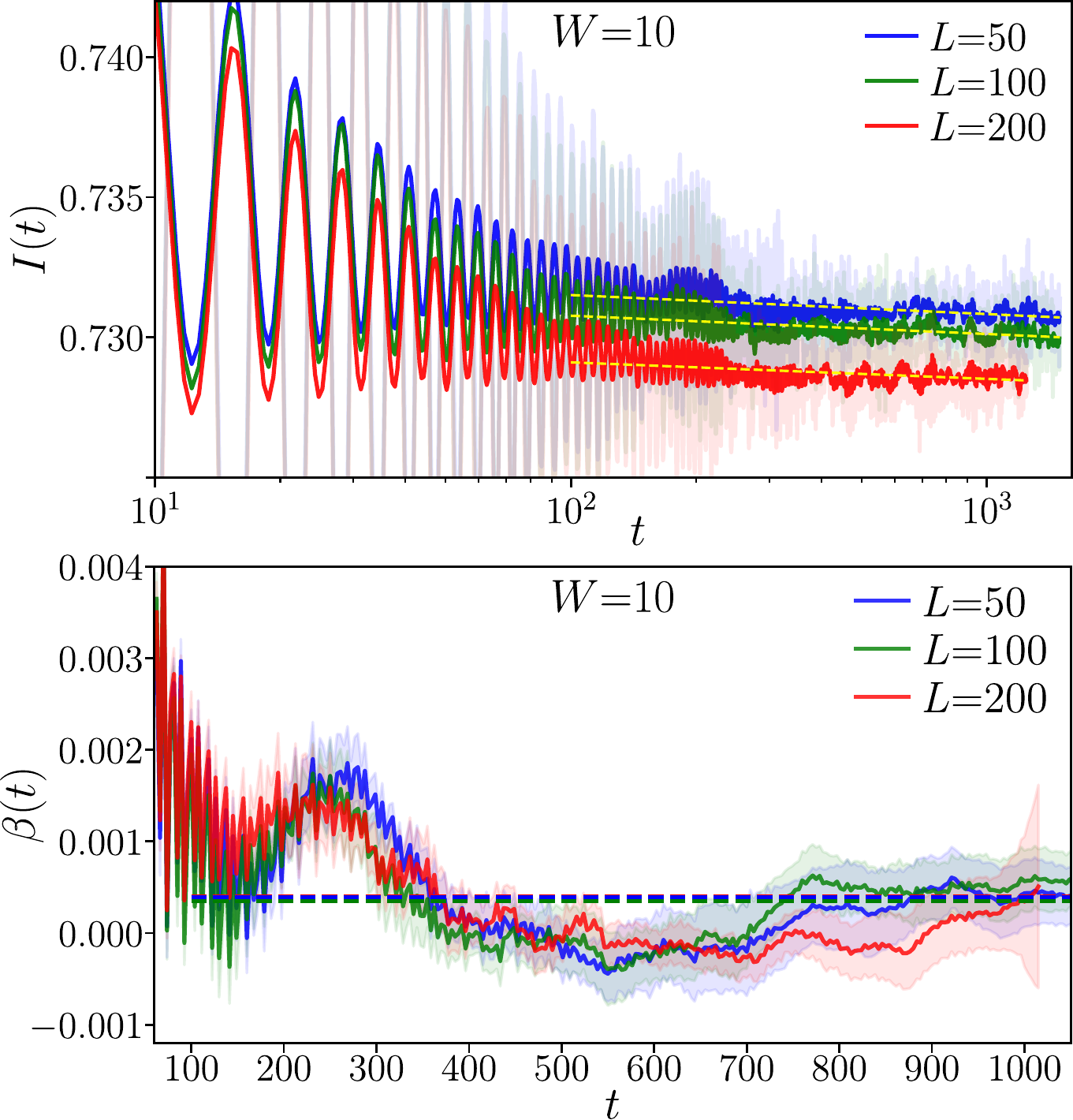}
  \caption{
Time evolution of imbalance for $W=10$, denotation the same as in Fig.~\ref{figRDW8}. Details of the simulations and fits given in Tab.~\ref{tab:table2}. 
 }
  \label{figRDW10}
\end{figure}  
  
  Time evolution of the imbalance $I(t)$, as well as the flowing $\beta(t)$ function are shown in Fig.~\ref{figRDW10}. 
\begin{table}
\caption{\label{tab:table2}
Details of numerical simulations for $W=10$, denotations the same as in Tab.~\ref{tab:table1}.
}
\begin{ruledtabular}
\begin{tabular}{lcccc}
{ } &
\textrm{$t_{\mathrm{max} }$}&
\textrm{$\chi$}&
\multicolumn{1}{c}{$n_{\mathrm{ real } }$ }&
\textrm{$\overline \beta$}\\
\colrule
$L$=50  & 1500 & 128 & 4000 &  $(3.93\pm0.82)\cdot 10^{-4} $\\
$L$=100  & 1500 & 128 & 2000 & $(3.60\pm0.53)\cdot 10^{-4}$\\
$L$=200  & 1200 & 160 & 1000 & $(3.50\pm0.87)\cdot 10^{-4}$  \\ \hline
$L$=50  & 5000 & 192 & 2000 &  $(3.08\pm0.51) \cdot 10^{-4}$ \\
\end{tabular}
\end{ruledtabular}
\end{table}
While the decay of imbalance clearly slowed down considerably,  as reflected by the values of the exponent $\overline \beta$ shown in Tab.~\ref{tab:table2}, upon the increase of disorder strength from $W=8$ to $W=10$, the system size dependence of $\overline \beta$ remains the same: the values of $\overline \beta$ are, within the estimated error bars, {similar} for $L=50,100,200$, clearly not satisfying the criterion \ref{critA}. The flowing $\beta(t)$ function, shown in the bottom panel of Fig.~\ref{figRDW10} indicates that the decay of imbalance is relatively fast around $t \approx 200$ and then slows down considerably at $t\approx 500$ for which the value of $\beta(t)$ is vanishing.  However, around $t\approx 800$ the flowing $\beta(t)$ function acquires again the value similar to $\overline \beta$ and the decay of imbalance persists and the criterion \ref{critB} is not met.

\begin{figure}
  \centering
\includegraphics[width=0.99\linewidth]{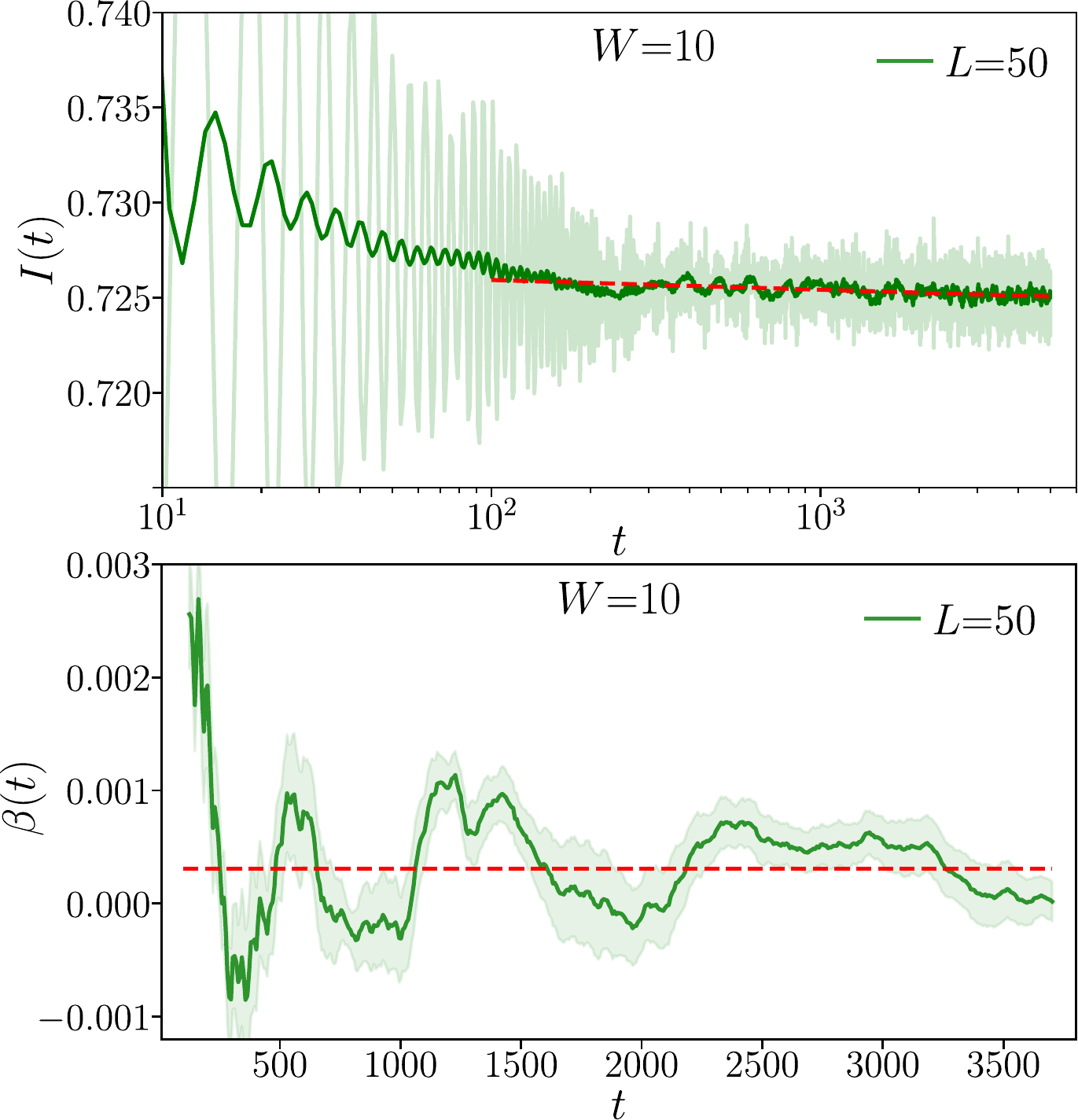}
  \caption{Time evolution of imbalance for $W=10$ in extended time interval, denotation the same as in Fig.~\ref{figRDW8}. Details of the simulations and fits given in Tab.~\ref{tab:table2}.
  }
  \label{figRDW10_long}
\end{figure}

To make sure that our conclusions for $W=10$ are valid, we increased the maximal time reached in our simulations to $t_{\mathrm{max} }=5000$ for system size $L=50$, the results are presented in Fig.~\ref{figRDW10_long}.
We indeed observe that the slow decay of imbalance $I(t)$ persists up to the longest time achieved in our simulation. This is exemplified by the power-law fit $I(t) \sim t^{- \overline \beta} $ that accurately matches the decay of imbalance in the whole interval $t\in[100,5000]$, with the exponent $\overline \beta$ close to the values obtained for the shorter time intervals, see Tab.~\ref{tab:table2}. Moreover, the flowing $\beta(t)$ function oscillates around the value $\overline \beta$ in the whole interval of available times. We see no signs of the slow down of decay of $I(t)$, which leads us to conclude that the criterion \ref{critB} is not fulfilled for $W=10$.

In conclusion, we found no clear signatures of the MBL phase in results presented in this sections, even though we considered significantly larger times and disorder strengths than in earlier studies \cite{Doggen18, Chanda19}. One immediate question is whether we can go even further in the attempts to observe MBL phase and consider larger disorder strength $W$ and bigger maximal time $t_{\mathrm{max}}$. The factor that limits such a continuation most severely is the slow-down of decay of $I(t)$ with $W$. In order to observe in a statistically significant way a decay of $I(t)$ at larger $W$ the increase of $t_{\mathrm{max}}$ should be coupled with an increase of the number of disorder realizations $n_{\mathrm{ real } }$. This considerably increases the resources needed for such numerical simulations. The same considerations apply to  experiments with quantum many-body systems which are limited by a finite coherence time ({typically limited to at most 1000 tunneling times \cite{Scherg21} thus shorter than the times considered by us}) as well as resources needed to perform disorder averages.

\section{Time evolution of entanglement entropy}
\label{sec:enta}
{The time dependence of the entanglement entropy is one of the tools that may be used to identify the existence of MBL phase. While typically
in the deconfined systems the entanglement entropy grows linearly in time when the evolution is started from the low entanglement, e.g. separable state, in MBL one expects a logarithmic entanglement entropy growth \cite{Znidaric08,Bardarson12}. It is, therefore, instructive to study 
the entropy growth in our case} 
 in the regime of large disorder strengths and long times probed in our numerical simulations.
Since the Hamiltonian \eqref{eq: XXZ} conserves the total magnetization $\sum_{i=1}^{L} S^z_i$, the entanglement entropy $S$ of subsystem $A$ consisting of lattice sites $1,\ldots L/2$ can be written as a sum $S(t)=S_n(t)+S_c(t)$, where {$S_n(t)$} is the number entropy and {$S_c$ denotes} the configurational entropy \cite{Schuch04,Schuch04b,Donnelly12,Turkeshi20,Lukin19,Sierant19c,Yao21}.
The number entropy is given by
\begin{equation}   
 S_n(t) = - \sum_{n} p(n) \ln p(n),
\end{equation}
where  $p(n)$ is the probability that $\sum_{i=1}^{L/2} S^z_i$ is equal to $n$. (We note that $\sum_{i=1}^{L/2} S^z_i$ is proportional to the total number of spinless fermions in subsystem $A$ after Jordan-Wigner transformation of \eqref{eq: XXZ}, explaining the term ``number entropy''.)
The configurational entropy is given by
\begin{equation}   
 S_c(t) = - \sum_{n} p(n) \mathrm{Tr}[\rho(n) \ln  \rho (n)],
\end{equation}
where $\rho(n)$ is the block of the reduced density matrix in sector with $\sum_{i=1}^{L/2} S^z_i=n$.

\begin{figure}
  \centering
\includegraphics[width=0.95\linewidth]{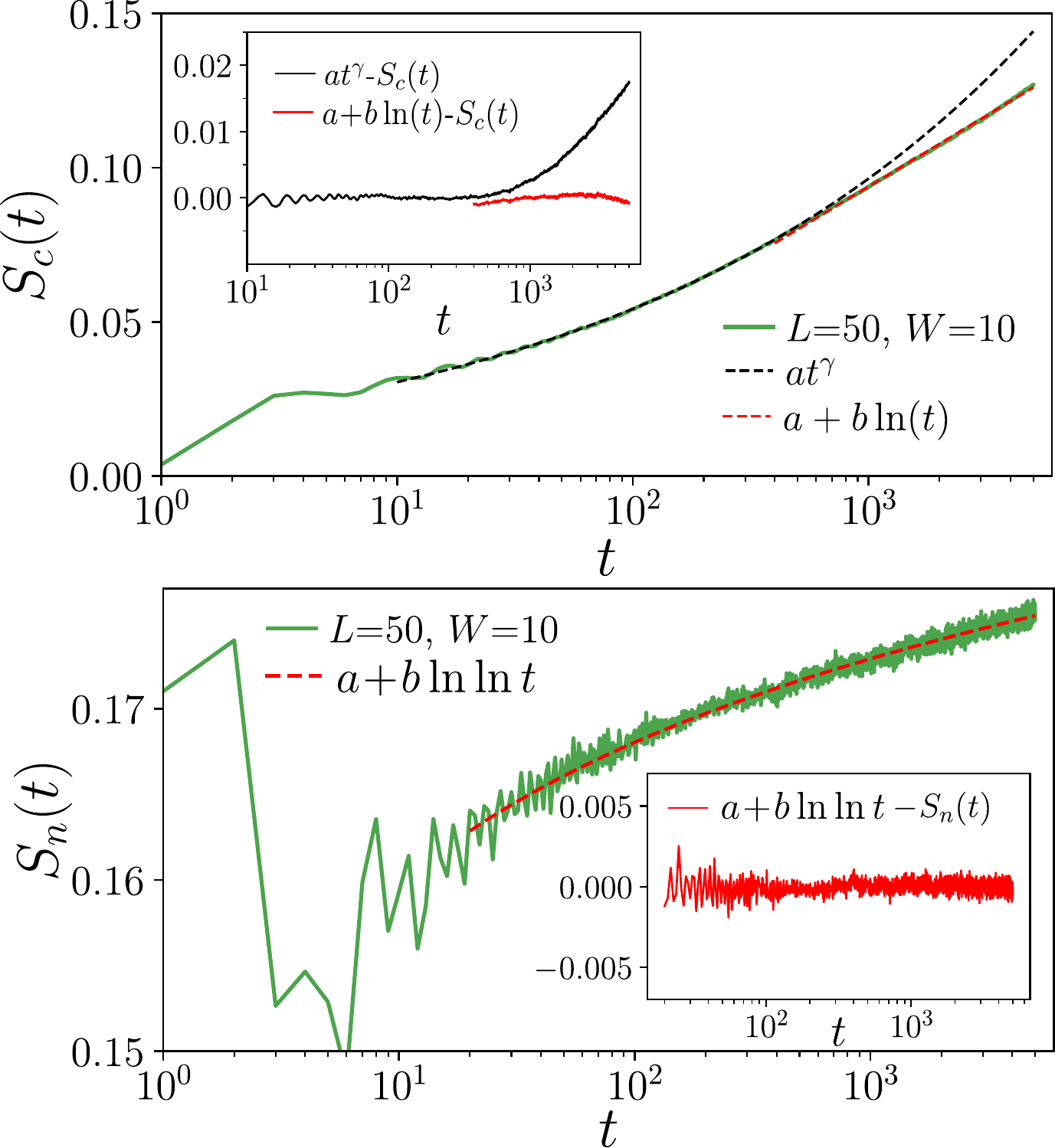}
  \caption{Time evolution of entanglement entropy for $L=50$ and $W=10$. Top: configuration entanglement entropy $S_c(t)$ denoted by solid line, dashed lines denote power-law and logarithmic fits $f(t)$. The inset shows the residual $f(t)-S_c(t)$. Bottom: The number entanglement entropy $S_n(t)$ is denoted by solid line, dashed line denotes a double-logarithmic fit $f_2(t)$. The inset shows the residual $f_2(t)-S_n(t)$.
  }
  \label{figEnt}
\end{figure}  
Our results for the entanglement entropies $S_n(t)$ and $S_c(t)$ are shown in Fig.~\ref{figEnt}. The configurational entropy $S_{c}(t)$ is expected to grow logarithmically in time \cite{Znidaric08, Bardarson12} in the MBL regime. We observe that after an initial transient at times $t \lessapprox 10$, the growth of $S_c(t)$ is well described by a power-law $S_c(t) \propto t^{\gamma}$ with $\gamma=0.250(2)$ in the interval $t\in[10,600]$. This behavior resembles the time dynamics of entanglement entropy observed in the ergodic regime at moderate values of disorder $W\approx2.5$ \cite{Luitz16}. However, at longer times, the increase of $S_c(t)$ slows down and  
is well fitted by $S_c(t) = a+b\ln t$ with $a=-0.04437(7)$ and $b=0.02001(9)$ for $t\in[400,5000]$ in agreement with expectations for the MBL regime. The growth of the number entropy is significantly slower, and is very well fitted by a double logarithmic formula $S_n(t) = a+b\ln \ln t$ with $a=0.1496(6)$ and $b=0.0120(3)$ in a wide regime of times $t\in[20,5000]$. This confirms the prediction of  \cite{Kiefer20, Kiefer21} for {the}  significantly larger system size and disorder strength than tested before.
 
In conclusion, the slow decay of imbalance observed in Sec.~\ref{sec:imba} is accompanied by a logarithmic increase of the configurational entanglement entropy $S_c(t)$ and a double logarithmic growth of the number entropy $S_n(t)$. Those quantities provide a complementary to the imbalance insight into the dynamics of the slow delocalization of the system. At the same time, they do not allow for an observation of the MBL phase in the fashion similar to the imbalance. For a localized system, one expects a saturation of $S_n(t)$  \cite{Luitz20}.
The upper limit, $S_n=\ln(3) {\approx 1.01}$, predicted in \cite{Luitz20}  is much higher than the values reached by a very slow double logarithmic growth of $S_n(t)$ observed in Fig.~\ref{figEnt}. Note also that a very recent study, \cite{Ghosh22}, instead of such a 
a slow double logarithmic growth predicts a power{-}law approach of ${S}_n$ to its {asymptotic} value {at $t \to \infty$}. This cannot be tested  {for the large system sizes $(L\geq50)$ considered by us since we are unable to determine} the asymptotic value {of $\lim_{t\to\infty} S_n(t)$.}

\section{Quasiperiodic systems}
\label{sec:imbaQP}

In this section we attempt at observation of MBL phase in dynamics of the system with QP potential{, defined in Sec.~\ref{sec:model}}. To that end we investigate the impact of the amplitude of QP potential  $W^{\mathrm{QP}}$ on time evolution of imbalance $I(t)$.

\begin{figure}
\includegraphics[width=0.99\linewidth]{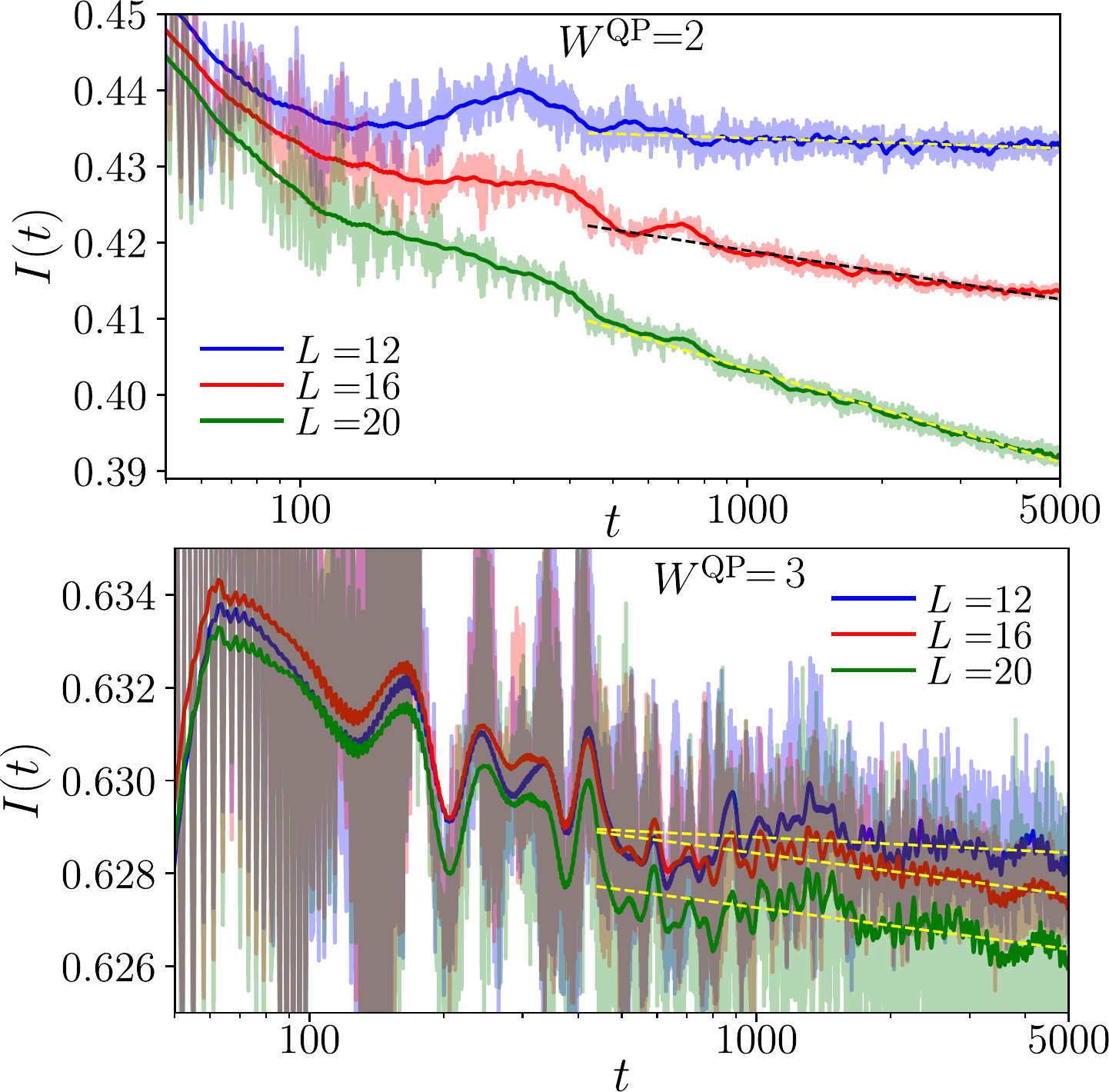}
  \caption{Time evolution of imbalance $I(t)$ for QP potential. Top: results for the amplitude of QP potential $W^{\mathrm{QP}}=2$ the shaded lines denote $I(t)$ whereas the solid lines denote a running overage of $I(t)$ over window $(t-25, t+25)$, dashed lines denote power-law fits $I(t) \sim t^{- \overline \beta} $ in time interval $t\in[500,5000]$. Bottom: the same for $W^{\mathrm{QP}}=3$. Details of simulations are given in Tab.~\ref{tab:table3}
  }
  \label{figQPD1}
\end{figure}  

The results for  $W^{\mathrm{QP}}=2, 3$ are shown in Fig.~\ref{figQPD1}. The behavior of $I(t)$ is qualitatively similar to the systems with random disorder: after an initial transient, the decay of imbalance is well fitted by a power-law  $I(t) \sim t^{- \overline \beta} $. The exponent $\overline \beta$ is clearly increasing with system size both for $W^{\mathrm{QP}}=2$ and $W^{\mathrm{QP}}=3$, as shown in Tab.~\ref{tab:table3},  suggesting that the system delocalizes in the thermodynamic limit at those values of $W^{\mathrm{QP}}$ and neither the criterion \ref{critA} nor \ref{critB} for observation of MBL phase is met.
 
  \begin{table}
\caption{\label{tab:table3}%
Details of numerical simulations QP potential, denotations the same as in Tab.~\ref{tab:table1}. The bond dimension $\chi$ is not displayed for calculation performed with the Chebyshev expansion of the evolution operator.
}
\begin{ruledtabular}
\begin{tabular}{lcccc}
{ } &
\textrm{$t_{\mathrm{max} }$}&
\textrm{$\chi$}&
\multicolumn{1}{c}{$n_{\mathrm{ real } }$ }&
\textrm{$\overline \beta$}\\
\colrule
$L$=12, $W^{\mathrm{QP}}$=2  & 5000 & - & $10^6$ &$ (1.8\pm0.2)\cdot 10^{-3} $\\
$L$=16, $W^{\mathrm{QP}}$=2  & 5000 & - & $10^5$ & $(9.5\pm0.2)\cdot 10^{-3}$\\
$L$=20, $W^{\mathrm{QP}}$=2  & 5000 & - & $5\cdot10^4$ & $(19.0\pm0.1)\cdot 10^{-3}$  \\ \hline
$L$=12, $W^{\mathrm{QP}}$=3  & 5000 & - & $10^6$ &$ (3.3\pm0.4)\cdot 10^{-4} $\\
$L$=16, $W^{\mathrm{QP}}$=3  & 5000 & - & $10^5$ & $(8.8\pm0.3)\cdot 10^{-4}$\\
$L$=20, $W^{\mathrm{QP}}$=3  & 5000 & - & $5\cdot10^4$ & $(8.9\pm0.6)\cdot 10^{-4}$  \\ \hline
$L$=12, $W^{\mathrm{QP}}$=4  & 5000 & - & $10^6$  &$ (2.1\pm0.4)\cdot 10^{-4} $\\
$L$=16, $W^{\mathrm{QP}}$=4  & 5000 & - & $10^5$ & $(2.8\pm0.3)\cdot 10^{-4}$\\
$L$=50, $W^{\mathrm{QP}}$=4  & 4000 & 128 &  & $(3.0\pm1.3)\cdot 10^{-4}$  \\ \hline
$L$=12, $W^{\mathrm{QP}}$=5  & 10000 & - & $10^6$ &$ (0.3\pm0.7)\cdot 10^{-4} $\\
$L$=16, $W^{\mathrm{QP}}$=5  & 10000 & - & $10^5$ & $(1.1\pm0.8)\cdot 10^{-4}$\\
$L$=50, $W^{\mathrm{QP}}$=5  & 4500 & 128 &  2000 & -  \\
$L$=100, $W^{\mathrm{QP}}$=5  & 3000 & 128 & 1000 & -  \\
$L$=200, $W^{\mathrm{QP}}$=5  & 2500 & 128 & 600 & - \\
\end{tabular}
\end{ruledtabular}
\end{table}

\begin{figure}
\includegraphics[width=0.99\linewidth]{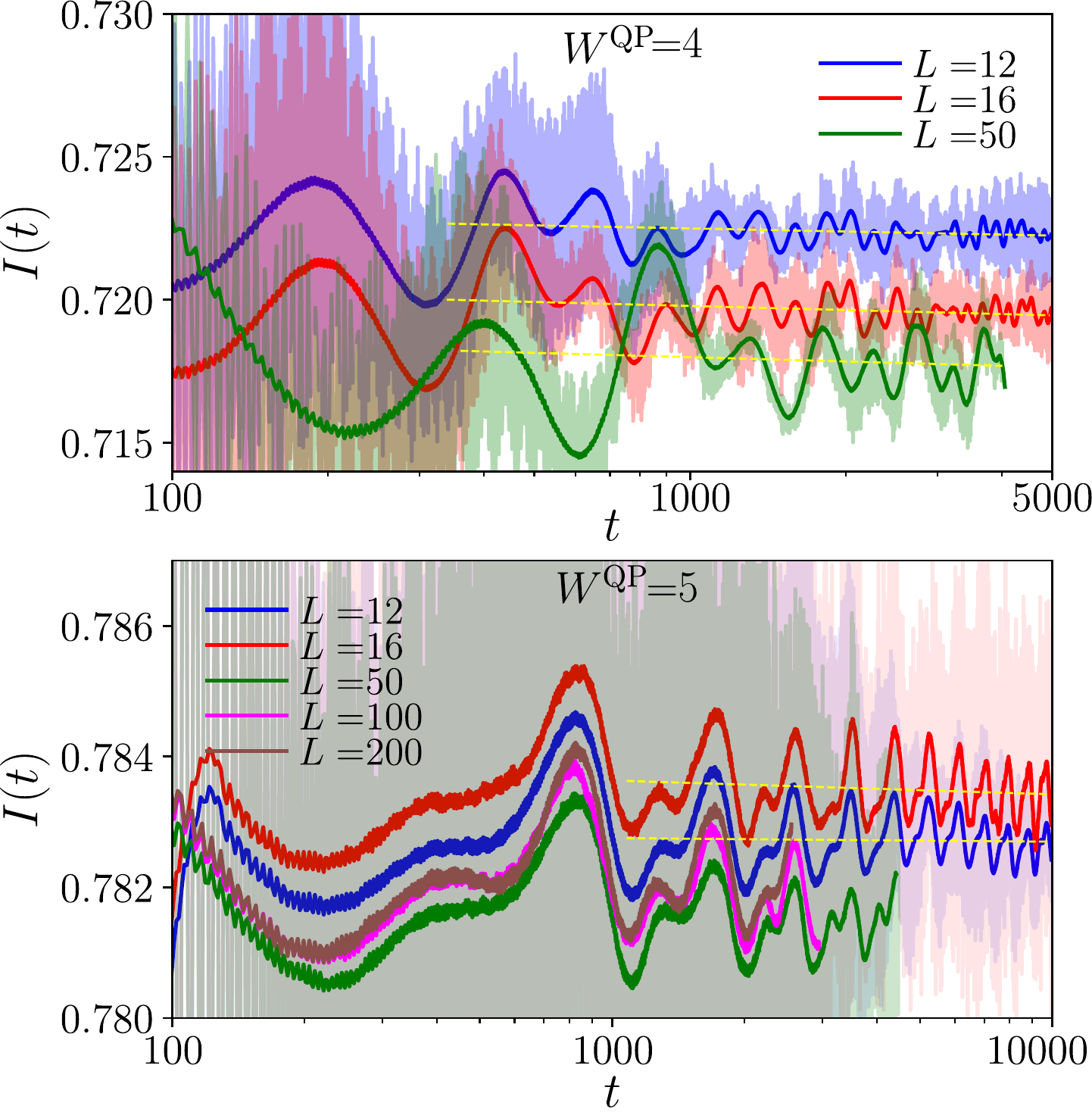}
  \caption{Time evolution of imbalance $I(t)$ for QP potential for $W^{\mathrm{QP}}=4,5$, denotation the same as in Fig.~\ref{figQPD1}. Details of the simulations and fits given in Tab.~\ref{tab:table3}.
  }
  \label{figQPD2}
\end{figure}  

The decay of imbalance $I(t)$ slows down considerably when the amplitude of the QP potential is increased to $W^{\mathrm{QP}}=4$ as shown in Fig.~\ref{figQPD2}. The exponents $\overline \beta$ governing the power-law decay of imbalance for $W^{\mathrm{QP}}=4$ are comparable to the exponents obtained for $W=10$ for the random disorder. However, the behavior of the running averages of $I(t)$ (shown by the solid lines  in Fig.~\ref{figQPD2}) is different: we observe significant oscillations around the fitted power-law decay. The pattern of those oscillations is not stable with increasing the system size, $L$.

This behavior changes qualitatively for $W^{\mathrm{QP}}=5$. For this amplitude of the QP potential we observe {an}  emergence of a pattern of oscillations of $I(t)$ at times $t \gtrapprox 200$ that remains the same when the system size is increased {from} $L=12$ to $L=200$. This is the first case for which we observe that the increase of the system size does not enhance its delocalization. Instead, this result shows that the dynamics of a small system comprised of  $L=12$ sites is reproduced in the bulk of the large system of $L=200$ sites. Such a behavior  suggests that the system remains MBL in the thermodynamic limit at $W^{\mathrm{QP}}=5${, although our approach is inherently limited to dynamics at finite times and cannot give a definite answer about the fate of the system at $t\rightarrow \infty$}. 

Two remarks are in order. Firstly, the values of the running average of $I(t)$ are not changing monotonically with $L$: the curve for $L=16$ is on the top whereas {that} for $L=50$ on the bottom. This is caused by the statistical fluctuations associated with the finite number of disorder realizations $n_{\mathrm{ real } }$ as well as by the erratic changes of $2\pi k L$ modulo $2 \pi$ with $L$ that  determine the number of full periods of the QP potential in the  {whole} chain. Secondly, the emergent pattern of oscillations of $I(t)$ prevents us from  determining whether the imbalance $I(t)$ slowly decays in time. Performing a power-law fit in the interval $t\in[1000,10000]$ we {have} found  non-vanishing values of $\overline \beta$ as shown in Tab.~\ref{tab:table3}. However, $\overline \beta$ changes significantly when the interval in which the fit is performed changes. 
{This shows that the criteria \ref{critA} and \ref{critB} are effectively inapplicable to the dynamics of imbalance in QP potential. }

{We refer the reader to Appendix~\ref{app:osc} for further numerical studies of the imbalance in QP potential where we show that the persistent oscillations tend to decay for {even} larger values of the  amplitude {$W^{QP}$.} We also show there that the character of the oscillations depends on the  parameter $k$ {which determines the quasiperiodicity of the potential} by {simulating} the dynamics also for $k=\sqrt{2}/2$.}

\begin{figure}
\includegraphics[width=0.99\linewidth]{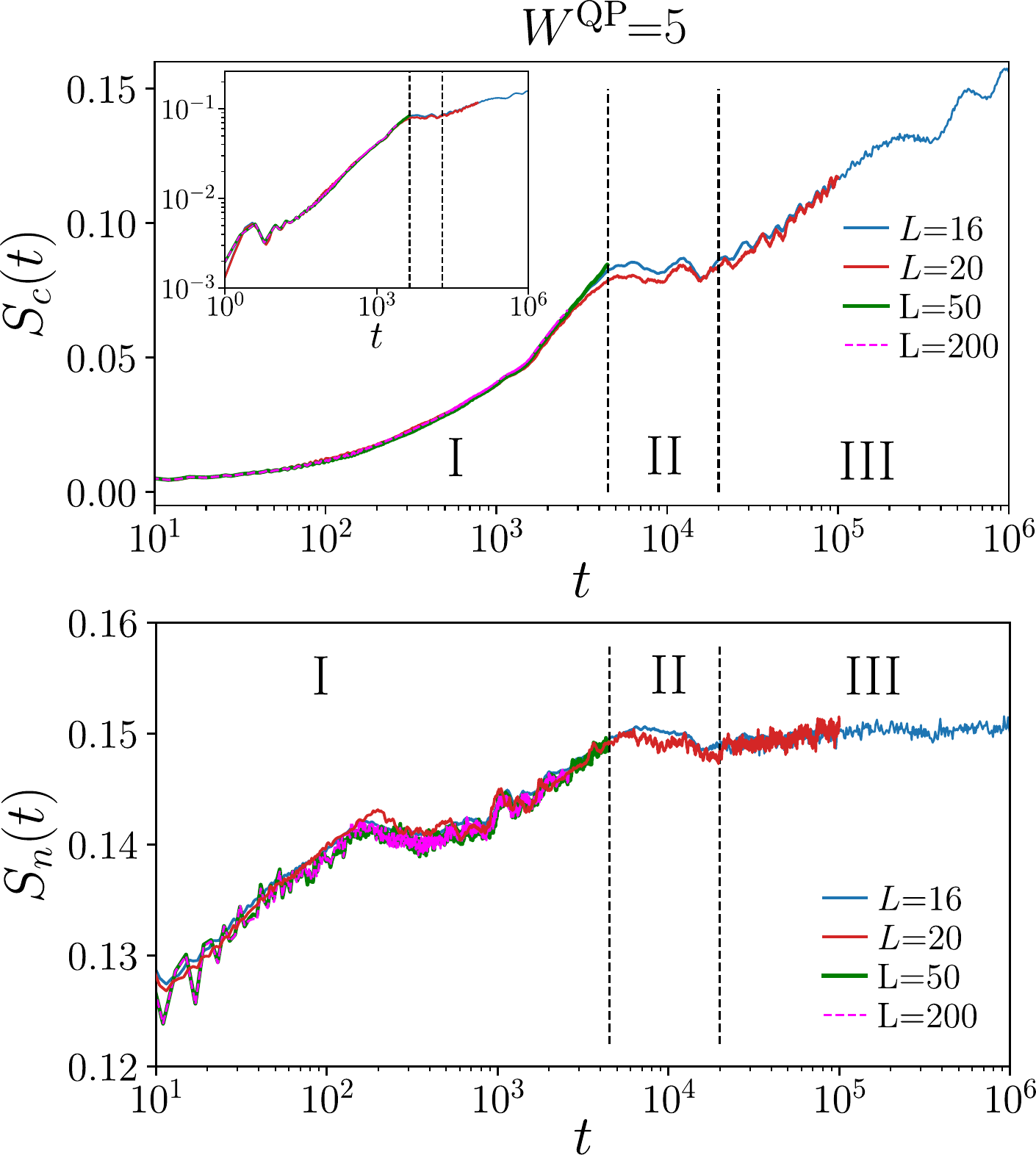}
  \caption{ {Time evolution of entanglement entropy for QP potential with $W^{QP}=5$. Top: configuration entanglement entropy $S_c(t)$ for system sizes $L=16,20,50,200$. The inset shows the same but on log-log scale. The dashed lines divide the time into intervals I, II, III (see text). Bottom: The corresponding number entanglement entropy $S_n(t)$.}
  }
  \label{figQPDent}
\end{figure}  
 
{To  explore the dynamics in the QP potential from a different perspective, we calculate the number, $S_n(t)$, and the configurational, $S_c(t)$, entropies for $W^{QP}=5$. The results are shown in Fig.~\ref{figQPDent}. Rather than observing an anticipated monotonic increase of the entanglement entropies, we may distinguish three time intervals (A,B,C) in the time dependence of $S_{c,n}(t)$. In the interval $I$, for $t\lesssim 4500$, we observe an algebraic in time increase of $S_c(t)$ (compare the inset in the top panel in Fig.~\ref{figQPDent} drawn in the log-log scale). This behavior is accompanied by a slow increase of the number entropy $S_n(t)$ which initially follows a logarithmic growth, saturates around $t\approx 300$ and then again seems to follow a logarithmic growth. In the interval I, the results for the small  $L=16,20$ and 
 large system sizes $L=50,200$ practically overlap both for $S_c(t)$ and $S_n(t)$. This is
  another property suggesting the locality of the dynamics at $W^{QP}=5$. 
 We observe for $L=16, 20$ that the behavior of $S_c(t)$ changes qualitatively at larger times: $S_c(t)$ is approximately constant in the region II ($4500<t<20000$) and grows logarithmically in time in the region III ($t>20000$). Both for $S_c(t)$ as well as for $S_{n}(t)$ the results for $L=16$ and $L=20$ are practically overlapping in the time intervals II and III. Unfortunately, the regimes II and III are inaccessible in TDVP calculations for large system sizes. This prevents us from deciding whether the initial power-law growth of $S_c(t)$ is continued in the large time limit for large system sizes (leading to a slow approach towards ergodicity) or whether the features of the entanglement growth at $L=20$ are consistent with the  behavior system for $L\rightarrow \infty$ (leading to a stable MBL phase).
 
}

\section{Conclusions}
\label{sec:conc}

In this work we have addressed the problem of {a possible experimental} observation of MBL. The presence of interactions gives rise to a slow dynamics towards equilibrium in strongly disordered systems. This lead{s} us to argue that an observation of even a very slow decay of correlation functions in a finite interval of time is insufficient {to claim an unambiguous observation of} MBL. 

For relatively small systems comprising of less than $L=20$ lattice sites, we calculated time dynamics beyond the Heisenberg time which allowed us to extrapolate the results to the infinite time limit. Building on intuitions obtained in that way, we formulated the criteria \ref{critA} and \ref{critB} {for an} observation of {the} MBL phase. The criterion \ref{critA} requires a slowdown of the decay of density correlation functions as $L^{-1}$ when the system size $L$ is increased. The criterion \ref{critB} demands a saturation of correlation functions {beyond a certain time scale} in a system size independent manner. We would like to emphasize that these criteria are neither sufficient nor necessary conditions to prove that a system is MBL. Rather, we perceive the criteria \ref{critA} and \ref{critB} as hints of whether the dynamics of a given system breaks the ergodicity or not.

Performing large scale tensor network simulations of time evolution of disordered XXZ spin chain of up to $L=200$ sites we did not find a regime of parameters in which the criterion \ref{critA} or \ref{critB} for observation of MBL would be satisfied. For considered disorder strengths we always encountered the slow but persistent decay of imbalance hinting at a slow approach of {the} system towards {the eventually delocalized future}. This conclusion was obtained even though we focused on the regime of disorder strengths lying significantly above the current estimates of the critical disorder strength for transition to MBL phase and pushed the maximal time reached in our simulations to few thousands tunneling times.
In that respect, our results are  consistent with the  nonexistence of MBL phase in the thermodynamic limit, see also \cite{Suntajs19,Sels21}.
We also revisited the dynamics of {the} entanglement entropy confirming the logarithmic growth of {its} configurational {part}  and the double logarithmic increase of the number entropy in the regime of long times and large system sizes confirming predictions of \cite{Kiefer20, Kiefer21}.

Finally, we investigated {the} time evolution of QP systems. The dynamics of quasiperiodic system is very much similar to random system at intermediate values of the amplitude $W^{\mathrm{QP}}$, with a slow, power-law like decay of imbalance. However, for a stronger QP potential, at $W^{\mathrm{QP}}=5$, we demonstrated an emergence of a pattern of oscillations in  the imbalance $I(t)$. {This pattern remains} stable with the increase of {the} system size. This qualitatively different behavior of the imbalance in a striking fashion shows that the dynamics of QP systems at sufficiently large potential strengths becomes local. While we were eventually not able to fully exclude the decay of the imbalance in the infinite time limit, the result for QP systems appears to be not far from being sufficient to claim an observation of MBL phase.   In any case, our results show that the asymptotic properties of transition to MBL phase may be probed more easily in QP systems (see \cite{Aramthottil21} for the analysis of QP system from the spectral perspective).

We would like to stress that our results, especially for disordered systems, do not  exclude the existence of a stable MBL phase. Rather, they provide lower bounds on time scales and disorder strengths required to observe the freezing of system dynamics in the long time limit that defines the MBL phase. 
Those lower bounds are relevant both for future numerical simulations of disordered systems as well as for experiments with quantum simulators.

 \begin{acknowledgments}
This work would not be possible without the help of Titas Chanda who provided us with his tensor network codes and generously helped with their
implementation. Thank you, Titas! We are also indebted to Anatoly Polkovnikov and Dries Sels for discussions as well as to Elmer V. H. Doggen for questions regarding our error analysis.
The numerical computations have been possible thanks to  the support of  PL-Grid Infrastructure.
The TDVP simulations have been performed using ITensor library (\url{https://itensor.org}).
This research has been supported by 
 National Science Centre (Poland) under project 2019/35/B/ST2/00034 (J.Z.)
 \end{acknowledgments}

\appendix

\section{Tests on the numerical accuracy of the presented results }
\label{app:num}
\
The results presented in the main text are obtained using different numerical techniques that will be described in detail below. {We also provide details of the numerical method for used for the non-interacting system.}

There are two types of errors in our results. The first is the statistical error which arises due to fluctuation of results from one disorder realization to another at fixed parameters of the system. The resulting errors in the exponent $\overline \beta$ governing the decay of imbalance $I(t)$, as well as in the running $\beta(t)$ function are estimated by the bootstrap technique, i.e. by resampling over the disorder realizations. In the figures we plot the imbalance $I(t)$ as well as a running average of the imbalance. Importantly, however, in the fits that determine $\beta(t)$ and $\overline \beta$  we always use the full data for the imbalance $I(t)$.
The second type of uncertainties are the systematic errors that might occur when the numerical simulations are not fully converged. Those systematic errors are particularly relevant for TDVP results. Below, we describe numerical tests that confirm that the values $\chi$ used by us in the main text are sufficient for the results to be converged, i.e. independent of the value of the bond dimension $\chi$.  

\subsection{ {Chebyshev time propagation} }

For small system sizes ($L \leq 20$) we use Chebyshev propagation scheme as described in detail in \cite{Fehske08}. In a nutshell, this approach approximates the time evolution operator $ U( \Delta t) = \exp(-i H \Delta t)$ over time period $\Delta t$ as
\begin{equation}
 U( \Delta t) \approx \mathrm{e}^{-\mathrm{i}b \Delta t} \left( J_0(a \Delta t) + 2\sum_{k=1}^N (-i)^k J_k(a \Delta t) T_k \left( \mathcal{H} \right) \right),
 \label{eqcheby}
\end{equation}
where $a=(E_{\rm max} - E_{\rm min})/2$, $b=(E_{\rm max} + E_{\rm min})/2$ and $E_{\rm min}$ ($E_{\rm max}$) is the energy of the ground state (the highest excited eigenstate) of the Hamiltonian $H$. The Hamiltonian is rescaled to $\mathcal{H} = \frac{1}{a}(H-b)$ so that spectrum of $\mathcal{H}$ belongs to the $[-1,1]$ interval,
$J_k(t)$ is the Bessel function of the order $k$ and $T_k(x)$ is the Chebyshev polynomial of order $k$.
The order of expansion $N$ needed to assure convergence of the expansion \eqref{eqcheby} for a given time step $\Delta t $ is computed in the following way. We take a random normalized state $\ket{\psi_R}$, calculate the state $U( \Delta t) \ket{\psi_R}$ with a certain trial order of expansion $N_{tr}$ and compute its norm. If the norm of $U( \Delta t) \ket{\psi_R}$ deviates from unity by more than $10^{-13}$, we know that $N_{tr}$ needs to be increased; otherwise $N_{tr}$ is decreased. This allows us to perform a binary search for $N_{tr}$ in the interval $N_{tr} \in [5,5000]$ (the upper boundary is determined by the maximal time step $\Delta t$ and parameters of the model). The result of this binary search, $N^0_{tr}$, is then incremented by $20\%$, yielding the desired order $N = 1.2N^0_{tr}$. We calculate the order of expansion whenever the time step $\Delta t$ changes in our algorithm. To calculate the time evolution of an initial state $\ket{\psi(0)}$ we repeatedly apply \eqref{eqcheby} to obtain $\ket{\psi(\Delta t)}$, $\ket{\psi(2 \Delta t)}$,$\ldots$,$\ket{\psi(t_{\mathrm{max}})}$. We have tested this procedure for system sizes $L\leq 16$ comparing $\ket{\psi(t_{\mathrm{max}})}$ with state $\ket{\psi_{ED}(t_{\mathrm{max}})} = U( t_{\mathrm{max}} )\ket{\psi(0)}$ evolved using time evolution operator $U( t_{\mathrm{max}} )$ determined by means of the full exact diagonalization of the Hamiltonian $H$. For $t_{\mathrm{max}} = 10^5$ we checked that the norm
$|| \ket{\psi(t_{\mathrm{max}})} - \ket{ \psi_{ED}(t_{\mathrm{max}})} || $ is smaller than $10^{-10}$ in the whole parameter range considered in this work. The deviation from unity of the norm of the state propagated with the Chebyshev expansion: $1 - || \ket{\psi(t_{\mathrm{max}})}|| $ was smaller than $10^{-12}$ for all system sizes considered in this work. For $L\leq 16$ the corresponding deviations in the value of $C(t)$ function (as compared to $\ket{\psi_{ED}(t_{\mathrm{max}})}$) were smaller than $10^{-13}$.
 We note that the main advantage of the Chebyshev expansion is that it efficiently utilizes the sparse matrix structure of the Hamiltonian of the system. This is due to the fact that a single time propagation step $U( \Delta t )\ket{\psi(t)}$ reduces to $\mathcal{O}(N)$ matrix-vector products and a calculation of linear combinations of vectors.

 \subsection{Tensor network approaches }
Chebyshev propagation scheme is not effective for larger system sizes {since it operates on the quantum states expressed as vectors in the full Hilbert space that is exponentially large in system size}. In contrast, tensor network techniques parameterize only a fraction of the full Hilbert space, encoding the state of the system in a matrix product state (MPS). This allows us to investigate time evolution of systems larger than $L>25$. The tensor network techniques were
developed over the years starting from seminal works of Vidal \cite{Vidal03,Vidal04} and White \cite{Feiguin04}. The link between the two approaches was illuminated in \cite{Daley04}. Those schemes are known as time-dependent density matrix renormalization group techniques (tDMRG) or time evolving block decimation TEBD techniques. The important modification came with the variational approach leading to algorithms based on Time Dependent Variational Principle (TDVP) optimal for an assumed limitation of the Hilbert space \cite{Haegeman11, Koffel12, Haegeman16, Goto19}. 
The time evolution can be calculated effectively with tensor network approaches only when the bond dimension $\chi$ of the MPS is sufficiently large to encode the state of the system. This gives rise to an upper limit on the entanglement entropy in the state of the system for a given $\chi$. This, in turn, translates into maximal time $t_{\mathrm{max}}$ to which time evolution of the system can be accurately simulated with TDVP/tDMRG for a given bond dimension $\chi$. Calculations in our work rely on the fact that for disorder strengths $W=8-10$ the spreading of entanglement in the system is very slow which allows us to probe the time evolution at times equal to few thousand tunneling times. 
The TDVP algorithm for time evolution consists of two stages. In the first stage we use the so called 2-site TDVP which allows for an accurate estimation of the errors. They appear mainly due to the truncation of the Hilbert space via Schmidt decomposition between the sites. When a disregarded Schmidt weight exceeds $10^{-12}$ the Hilbert space is enlarged so in this stage the algorithm is practically exact until the bond dimension reaches the prescribed value $\chi$ at a given bond. At this stage we switch (at this bond) to 1-site TDVP algorithm, from this moment errors due to the Hilbert space truncation start to accumulate. This is a standard, well developed  strategy \cite{Chanda19, Chanda20m} which we follow in our work.

The detailed comparison of the performance of TEBD and TDVP algorithms  for the random-field XXZ chain, but for lower disorder amplitudes than in the present work, was performed in in our previous work  \cite{Chanda19}. It was shown, in particular, that the
 TEBD algorithm that is unconverged, i.e. the bond dimension is not sufficiently large to follow the time evolution of the state up to the requested time $t_{\mathrm{max}}$, spuriously indicates a stabilization of the imbalance $I(t)$ suggesting a localization in the system.  In contrast, unconverged TDVP has a tendency to show a delocalization in the system by overestimating the degree of decay of the imbalance $I(t)$. {An analogous behavior of TDVP was also observed in a different {disorder-free models} in \cite{Goto19}.} 
\begin{figure}
  \centering
\includegraphics[width=0.95\linewidth]{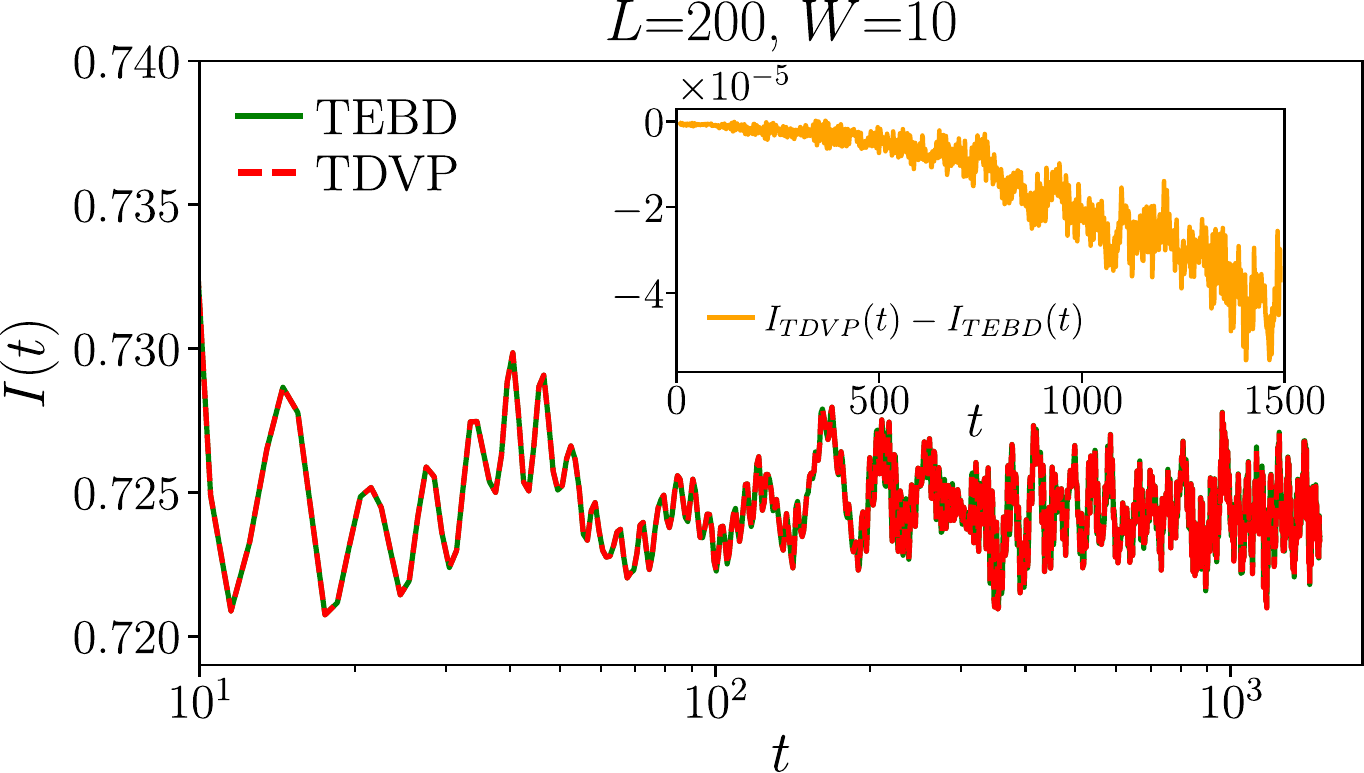}
\vspace{-0.2cm}
  \caption{Comparison of the imbalance $I(t)$ (averaged over times $[t-10,t+10]$) for system size $L=200$ and disorder strength $W=10$ obtained with TEBD and TDVP algorithms. The bond dimension is fixed as $\chi=128$ and the results are averaged over $24$ disorder realizations. The inset shows the difference between the imbalances $I(t)$ for TDVP and TEBD propagation schemes.
  }
  \label{figTEBDvTDVP}
\end{figure}  
\begin{figure}
  \centering
\includegraphics[width=0.95\linewidth]{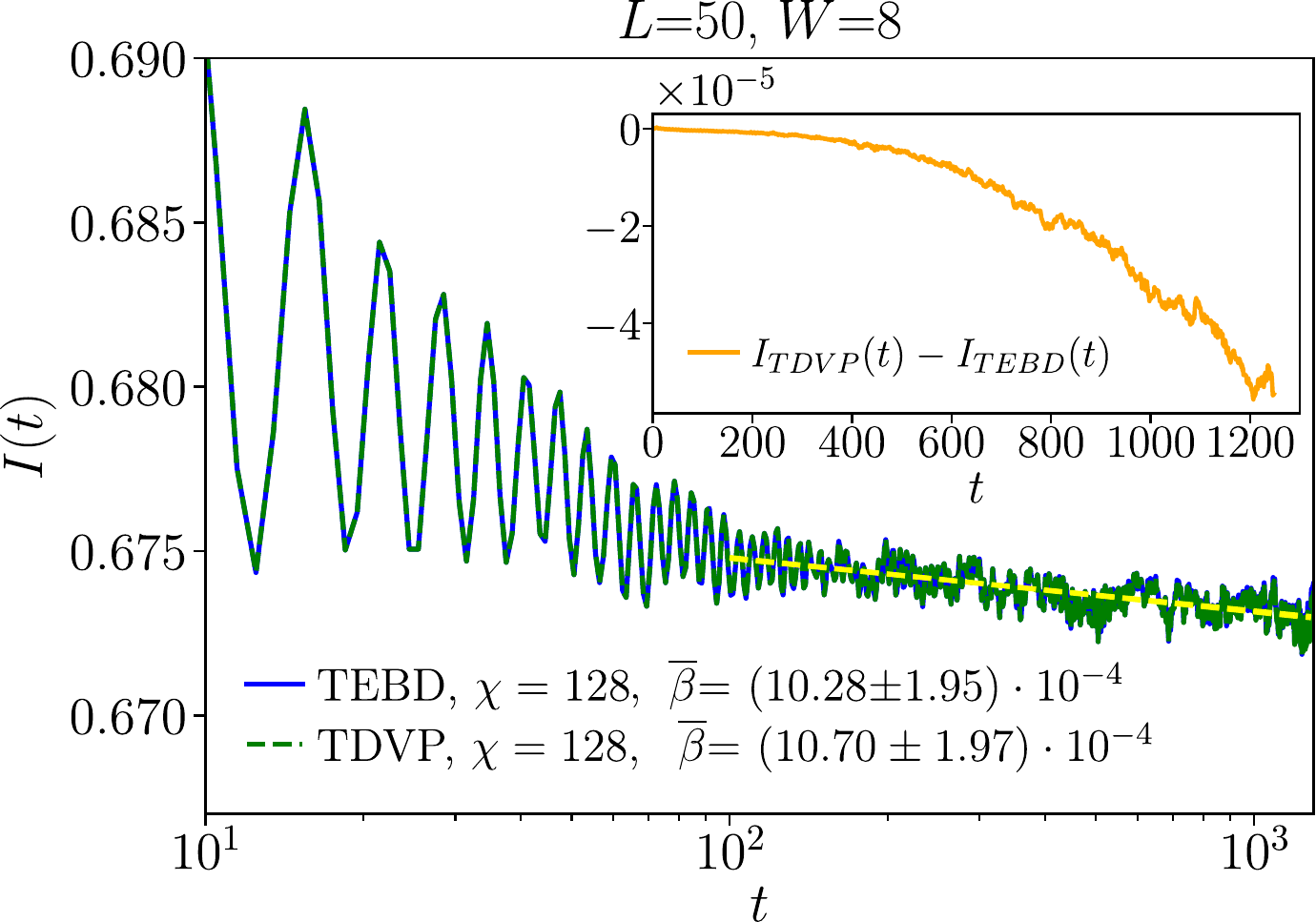}
\vspace{-0.2cm}
  \caption{{Comparison of the imbalance $I(t)$ (averaged over times $[t-10,t+10]$) for system size $L=50$ and disorder strength $W=8$ obtained with TEBD and TDVP algorithms. The bond dimension is fixed as $\chi=128$ and the results are averaged over $1000$ disorder realizations. The dashed line shows the fitted power-law decay of $I(t)$. The inset shows the difference between the imbalances $I(t)$ for TDVP and TEBD propagation schemes.}
  }
  \label{figTEBDvTDVP2}
\end{figure}  
This motivates us to compare results for the imbalance $I(t)$ obtained with TEBD and TDVP algorithms as shown 
 in Fig.~\ref{figTEBDvTDVP}. The results are averaged over 24 disorder realizations for $L=200$ and $W=10$. We observe that the agreement between
TDVP and TEBD results is excellent indicating the convergence for individual disorder realizations. {T}he difference between the curves at late times oscillates around ${ 4} \cdot 10^{-5}$. This small discrepancy can be compared with the total change of the value of imbalance $\Delta I = 7\cdot10^{-4}$ in the interval $t\in [100,1200]$ for $L=200$ (cf. Fig.~\ref{figRDW10}). The latter value is more than an order of magnitude larger than the discrepancy between TDVP and TEBD results. This suggests that the exponent of power-law decay $\overline \beta = (3.50\pm0.87)\cdot 10^{-4}$ for $L=200$ (see Tab.~\ref{tab:table2}) is accurately estimated.
 
While TEBD is faster ``per time step'' for such a large disorder amplitude ($W=10$) we must take very small time step $\Delta t=0.001$ for TEBD to obtain converged results. {T}he error of the approximate unitary evolution may be estimated by a relative energy change in TEBD algorithm {$[(\braket{ \psi_{TEBD}(t_{\mathrm{max}}) |H| \psi_{TEBD}(t_{\mathrm{max}})} -E_0)/E_0$ where $ \ket{\psi_{TEBD}(t)}$ is the state obtained in TEBD time evolution and $E_0 = \braket{\psi(0)|H| \psi(0)}$}].  It remains below $10^{-4}$  for even  the most unfavorable disorder realisation (for $\chi=128$). {At the same time,} the total accumulated error, equal to sum of squares of Schmidt coefficients disregarded in all time steps, {associated with} necessary truncations inherent to TEBD is below $10^{-5}$. 
As shown in the following, the resulting error is sufficiently small to obtain an accurate estimate of the exponent $\overline \beta$. The required small step makes, however, {the} application of TEBD scheme not practical.  For large disorder amplitudes and large time scales, it is more efficient to use TDVP.  It allows us to keep the time step at a reasonable value, $\Delta t=0.1$ ({the time scale is fixed by $J=1$ in \eqref{eq: XXZ}}). We have checked by decreasing the time step that the chosen value leads to accurate results. The agreement of TDVP results with the numerically exact results for the non-interacting case, shown in Fig.~\ref{figAnd} provides another test of the convergence  of our results with the time step.

{While the comparison for $L=200$ is carried out for 24 disorder realizations only, we supplement it with comparison for $W=8$ and $L=50$ carried out  for over a 1000 disorder realizations in Fig.~\ref{figTEBDvTDVP2}. Again the discrepancy between curves is small (of the order of ${ 4} \cdot 10^{-5}$ as for $L=200$ data in Fig.~\ref{figTEBDvTDVP}. Larger number of disorder realizations allows us to extract  reliably $\beta$ values from both simulations. They agree very well with each other indicating the agreement between both aogorithms used.}

\begin{figure}
  \centering
\includegraphics[width=0.95\linewidth]{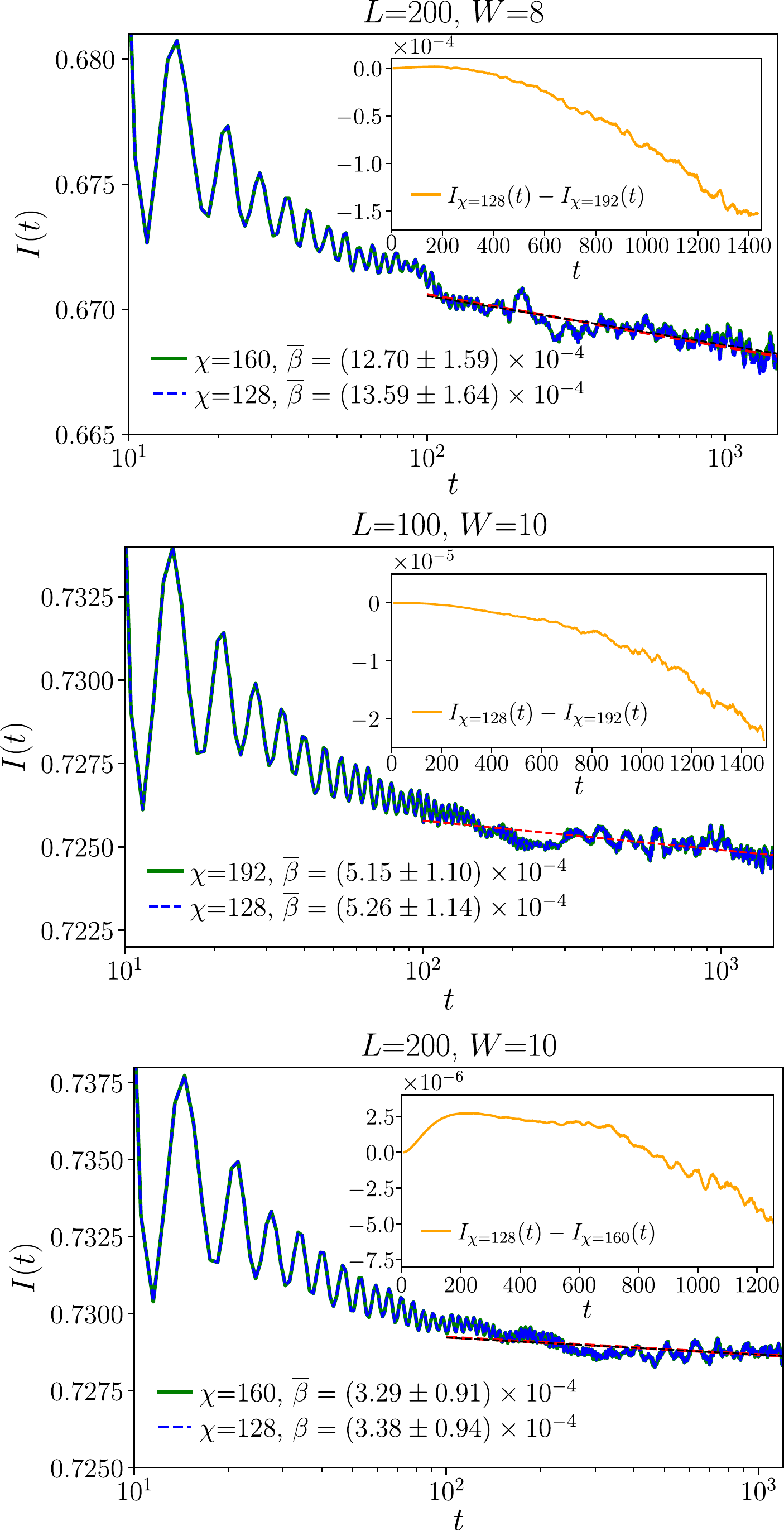}
\vspace{-0.2cm}
  \caption{The imbalance $I(t)$ (averaged over times $[t-10,t+10]$) for system size $L$ and disorder strength $W$ obtained with TDVP algorithm with bond dimension $\chi$, dashed lines show power-law fits $I(t) \sim t^{-\overline \beta}$ in time interval $t\in[100, t_{\mathrm{max}}]$ . Top panel: $L=200$, $W=8$, results averaged over 250 disorder realizations, $t_{\mathrm{max}}=1500$. Center panel: $L=100$, $W=10$, results averaged over 1000 disorder realizations,$t_{\mathrm{max}}=1500$. Bottom panel: $L=200$, $W=10$, results averaged over 984 disorder realizations, $t_{\mathrm{max}}=1200$. {The insets show the difference between the imbalances for the larger and the smaller value of $\chi$. }
  }
  \label{figTDVPcvg1}
\end{figure}  

\begin{figure}
  \centering
\includegraphics[width=0.95\linewidth]{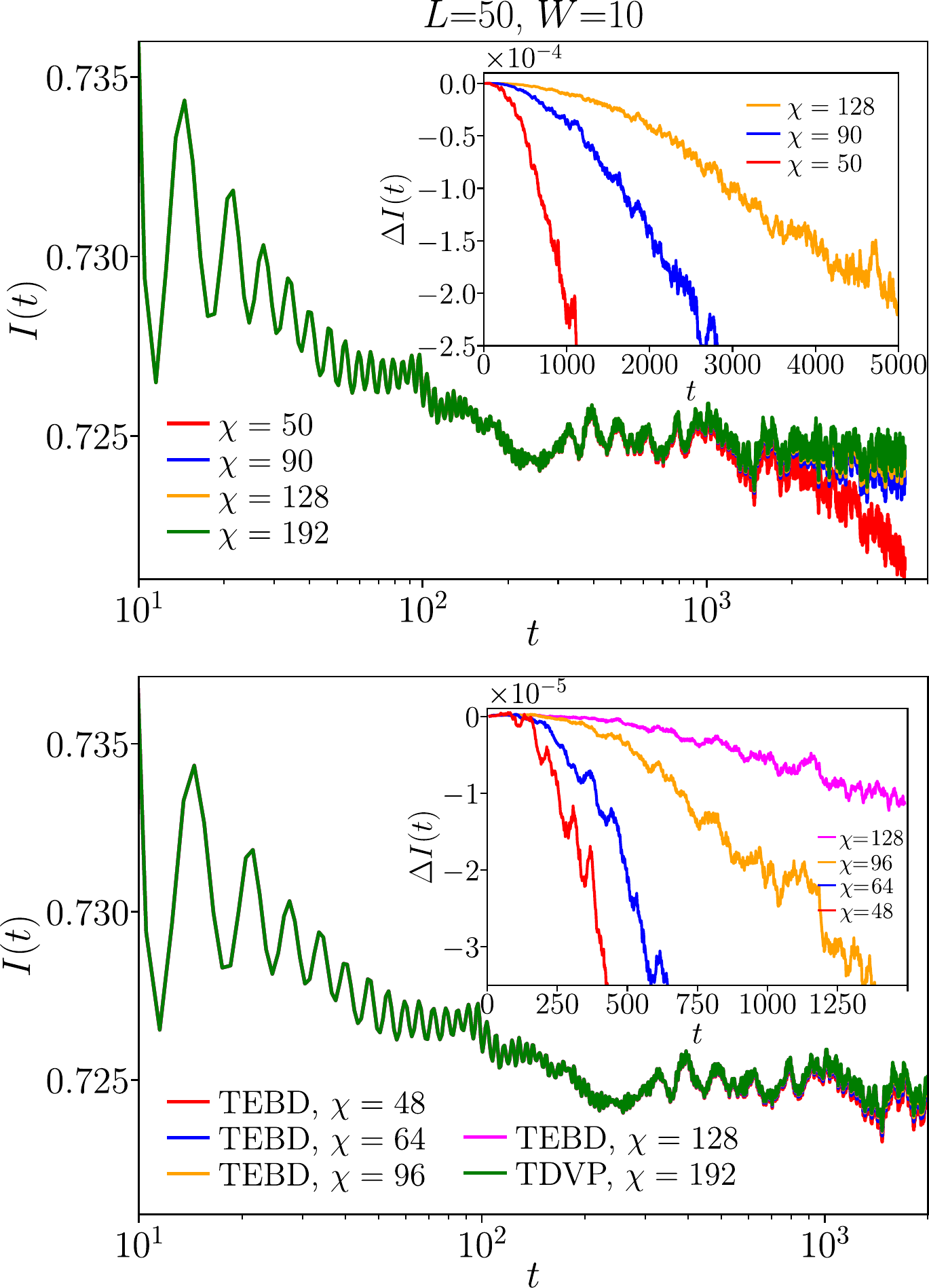}
  \caption{{Top panel:} Comparison of the imbalance $I(t)$ (averaged over times $[t-10,t+10]$) for system size $L=50$ and disorder strength $W=10$ obtained with TDVP algorithm with bond dimension $\chi = [50,90,128,192]$. The results are averaged over $1000$ disorder realizations. {The inset shows the difference $\Delta I(t)=I_{\chi}(t)-I_{\chi=192}(t)$ between the imbalance obtained with TDVP with the largest bond dimension $\chi=192$ and the imbalances obtained with $\chi=128,90,50$. Bottom panel: the same, but data for the smaller bond dimensions: $\chi=48, 64, 96, 128$ obtained with TEBD algorithm. Note the difference in the range of the horizontal axes of the two panels. }
  }
  \label{figTDVPcvg2}
\end{figure}  
\begin{figure}
  \centering
\includegraphics[width=0.8\linewidth]{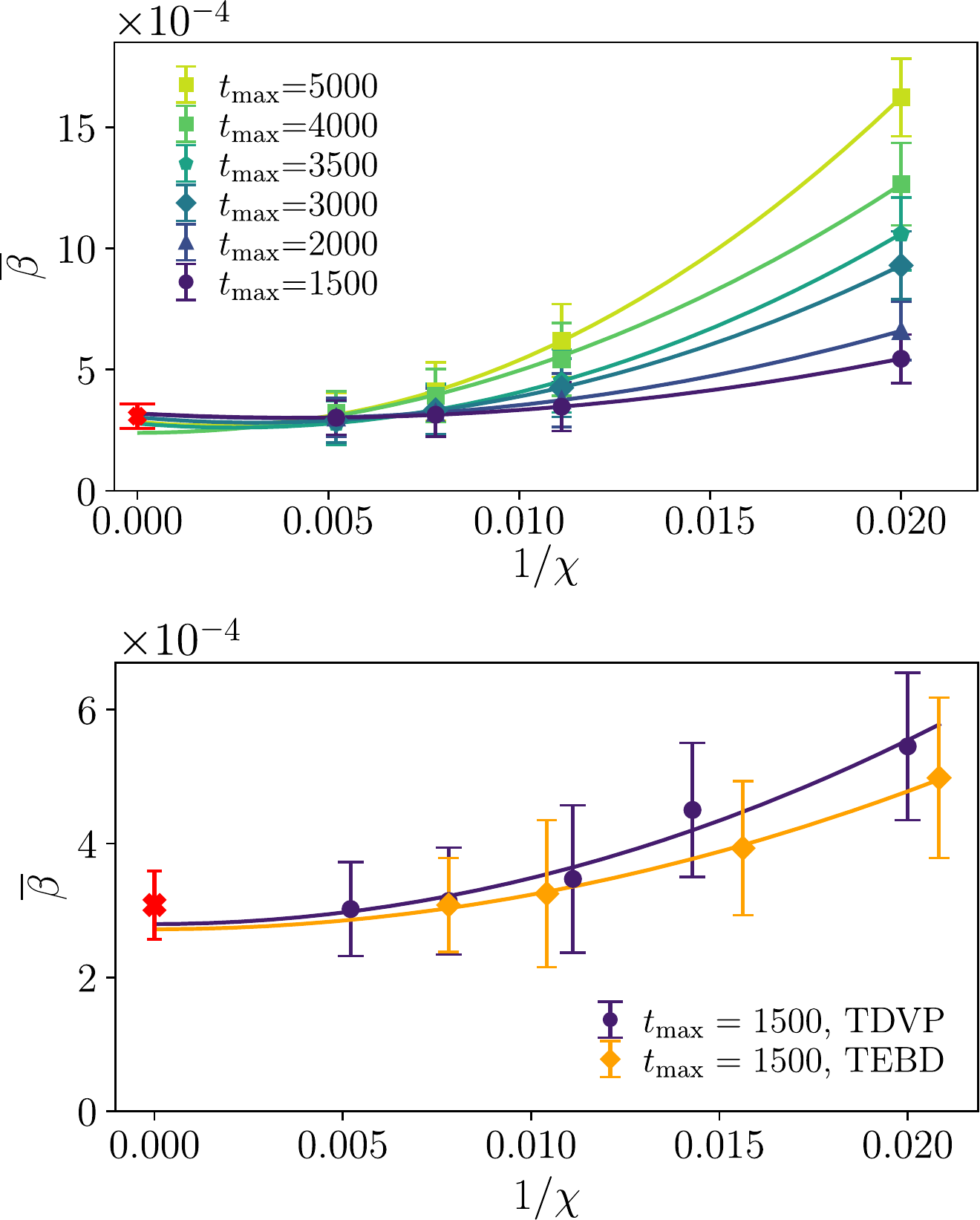}
  \caption{  {Top panel: Comparison of the values of the exponent $\overline \beta$ governing the decay of the imbalance $I(t)$ for disorder strength $W=10$ and system size $L=50$ obtained with TDVP propagation scheme. The fitting was performed in the interval $t\in[100, t_{\mathrm{max}}]$ and results are shown as a function of $1/\chi$, solid lines denote second order polynomial fits in $1/\chi$, the red point at $1/\chi=0$ shows the result  $\overline \beta = (3.08\pm0.51) \cdot 10^{-4}$ from Tab.~\ref{tab:table2}. Bottom panel: The exponent $\overline \beta$  as a function of $1/\chi$ for TDVP and TEBD algorithms (for $L=50$, $W=10$). The values of $\overline \beta$ presented in both are extracted from data shown in Fig.~\ref{figTDVPcvg2}. }}
  \label{figTDVPcvg2b}
\end{figure}  
 
The {above} comparison of TEBD and TDVP algorithms suggests a good convergence of our TDVP results. To further investigate the accuracy of the TDVP scheme, we compare results obtained for a varying bond dimension $\chi$. Fig.~\ref{figTDVPcvg1} summarizes our results. In each of the investigated cases we observe that the curves showing the imbalance $I(t)$ practically overlap for the both bond dimensions considered {(cf. the insets in Fig.~\ref{figTDVPcvg1})}. We observe that the exponents $\overline \beta$ govering the power-law decay of imbalance for smaller and larger $\chi$ are consistent with each other indicating a good convergence of the data with the bond dimension. At the same time, we observe that the $\overline \beta$ slightly decreases with the increase of the bond dimension $\chi$ in each of the analyzed cases. 

This dependence is further analyzed in {the top panel of} Fig.~\ref{figTDVPcvg2} in which we have supplemented the data for $\chi=128$ and $\chi=192$ with results for smaller bond dimensions $\chi=50,90$. Interestingly, the agreement of results for $\chi \geq 50$ up to time $t \approx 1000$ shows that already the results for $\chi=50$ are a good estimate of the imbalance $I(t)$ in that time interval at $W=10$. At larger times, the results for $\chi=50$ are unconverged and show a spurious signatures of delocalization in the system consistently with our expectations based on \cite{Chanda19}.
{The bottom panel of Fig.~\ref{figTDVPcvg2} compares the TDVP results for $\chi=192$ with the imbalance obtained with TEBD and bond dimensions $\chi=48, 64, 96, 128$. Contrary to the expectations from \cite{Chanda19}, we see that TEBD also indicates weaker and weaker decay of the imbalance $I(t)$ as the bond dimension $\chi$ is increased. The difference between the TDVP and TEBD results for the largest $\chi$ presented is no bigger than $ 4\cdot 10^{-5}$ indicating that both algorithms yield consistent estimates of $\overline \beta$, $\beta(t)$. }

To clarify the dependence of the results on the value of $\chi$, we plot the values of the exponent $\overline \beta$ as function of $1/\chi$  in {the top panel of Fig.~\ref{figTDVPcvg2b}}. The value of the exponent $\overline \beta$ decreases monotonously with the bond dimension $\chi$. The change in the value of $\beta$ when $\chi$ increases from $50$ to $192$ is the smallest for $t_{\mathrm{max}}=1500$ (indicating that smaller bond dimensions are needed to get converged results for $t<1500$), and increases with the increase of $t_{\mathrm{max}}$. Nevertheless, the extrapolations of $\overline \beta$ with a second order polynomial in $1/\chi$ give consistent results for all considered values of $t_{\mathrm{max}}$. Importantly, those extrapolations are in agreement with the result $\overline \beta = (3.08\pm0.51) \cdot 10^{-4}$ from Tab.~\ref{tab:table2}, confirming the convergence of our simulations with the bond dimension. {The bottom panel of Fig.~\ref{figTDVPcvg2b} shows a comparison of $\overline \beta$ for $t_{\mathrm{max}}=1500$ for TEBD and TDVP results. The values of $\overline \beta$ are nearly independent of $\chi$ for $\chi\gtrapprox 90$  confirming that both algorithms are very close to being converged at those bond dimensions for $t< t_{\mathrm{max}}=1500$. Finally, the extrapolation of those results to large $\chi$ limit yields the consistent values of $\overline \beta$ for both TEBD and TDVP in line with our message about the persistence of a slow decay of the imbalance even at the large disorder strength $W=10$. }

 \subsection{ {Time evolution for free fermions} }
 \label{app:free}
 
 Here, for completeness, we provide details of the standard (see \cite{Peschel09} and references therein) approach to time evolution of a system of non-interacting fermions used by us in Sec.~\ref{sec:nonint}. The Hamiltonian \eqref{eq: XXZ}, upon Jordan-Wigner transformation, becomes 
 \begin{equation}
 \hat{H}= 2J \sum_{i=1}^{L-1} \left( \hat{c}^{\dag}_i \hat{c}_{i+1} + \hat{c}^{\dag}_{i+1} \hat{c}_{i}+ \frac{\Delta}{2} \hat{n}_i \hat{n}_{i+1}  \right) + \sum_{i=1}^{L} h_i \hat{n}_i,
 \label{eq:ff}
\end{equation}
where $\hat{c}^{\dag}_i$ ($\hat{c}_i$) is creation (anihilation) operator of spinless fermion at site $i$, canonical anti-commutation relation $\{\hat{c}_i,\hat{c}^{\dag}_j \}= \delta_{ij}$ is fulfilled, and the number operator is given as $\hat{n}_i=\hat{c}^{\dag}_i\hat{c}_i$. For $\Delta=0$, the model \eqref{eq:ff} becomes non-interacting. Then, it can be written as a quadratic form of the fermionic operators
 \begin{equation}
 \hat{H}= \sum_{i,j=1}^L h_{ij} \hat{c}^{\dag}_i \hat{c}_j, 
 \label{eq:ff2}
\end{equation}
where we have introduced a $L\times L$ matrix $\mathbf{h}= ( h_{ij})$. Time dependence of the fermion anihilation operator is given by
\begin{equation}
 \hat{c}_i(t)= e^{i  \hat{H} t} \, \hat{c}_i e^{-i  \hat{H} t} =\sum_{j=1}^L (e^{-i \mathbf{h} t})_{ij} \hat{c}_j,
 \label{eq:ff3}
\end{equation}
where the second equality can be obtained from the Baker–Campbell–Hausdorff formula. Defining a  $L\times L$ correlation matrix 
\begin{equation}
 \mathbf{C}(t)= (\mathbf{C}(t))_{i,j}= \bra{\psi} \hat{c}^{\dag}_i(t)\hat{c}_i(t)  \ket{\psi},
 \label{eq:ff4}
\end{equation}
and using \eqref{eq:ff3}, we find that 
\begin{equation}
 \mathbf{C}(t)= e^{i \mathbf{h} t} \mathbf{C}(0) \,e^{-i \mathbf{h} t}.
 \label{eq:ff5}
\end{equation}
The correlation matrix $\mathbf{C}(0)$ at $t=0$ is determined by the initial state, and for the N\'eel state the only non-vanishing coefficients are $\mathbf{C}(0)_{2k,2k}=1$ where $k=1,\ldots,L/2$. The imbalance is given by
\begin{equation}
 I(t) = D  \sum_{i=1+l_0}^{L-l_0}   (-1)^{i} (\mathbf{C}(t))_{ii}, 
 \label{eq:ff6}
\end{equation}
where the constant $D$ assures that $I(0)=0$. Finally, to calculate entanglement entropy for a bipartition of the system into subsystems consisting of sites $1,\ldots,l_A$ and $l_A+1,\ldots L$, we calculate eigenvalues $\lambda_i$ of the submatrix $(\mathbf{C}(t))_{i,j=1}^{l_A}$ and compute the entanglement entropy as 
 \cite{Peschel03, Cheong04}
\begin{equation}
 S(t) = - \sum_{i=1}^{l_A}  [ \lambda_i \ln(\lambda_i) + (1-\lambda_i) \ln(1-\lambda_i)].
 \label{eq:iff7}
\end{equation}

The formulas \eqref{eq:ff6} and \eqref{eq:iff7} allow us to calculate the imbalance and entanglement entropy for the XXZ spin chain with $\Delta=0$ with numerical cost scaling as $L^3$.

\begin{figure}[h]
  \centering
\includegraphics[width=0.92\linewidth]{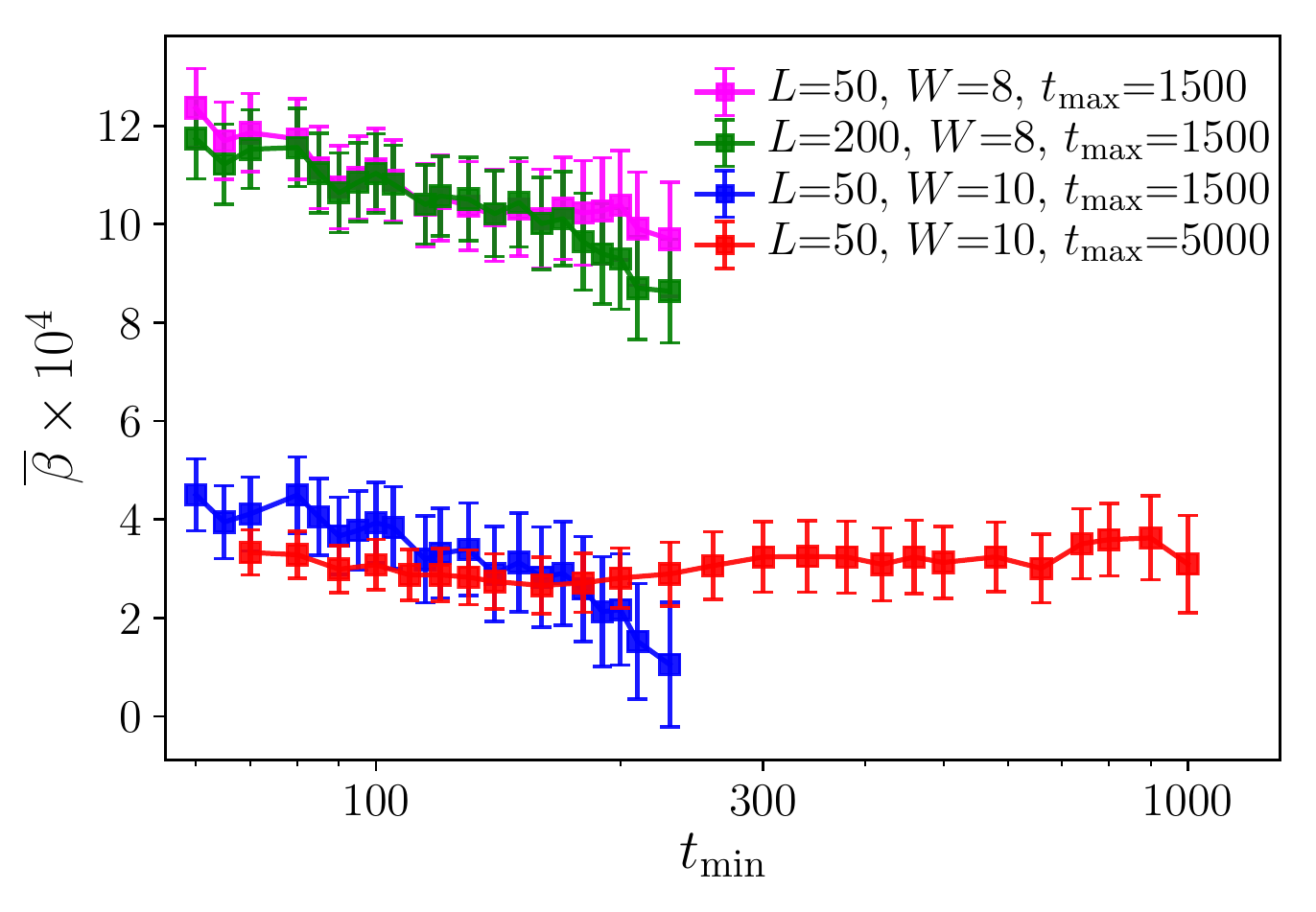}
\vspace{-0.4cm}
  \caption{The exponent $\overline \beta$ obtained from the fit $I(t) \sim t^{-\overline  \beta}$ in interval  $t \in [t_{\mathrm{min}}, t_{\mathrm{max}}]$ as a function of $t_{\mathrm{min}}$ for $W=8,10$ and system size $L=50,200$.
  }
  \label{figTmin}
  \end{figure}  
\section{Stability of the power-law fits to choice of time interval }
\label{app:tmin}

In the main text, the imbalance was fitted by a power-law decay: $I(t) \sim t^{-\overline  \beta}$ in interval $t \in [t_{\mathrm{min}}, t_{\mathrm{max}}]$, where $t_{\mathrm{min}}=100$ and the value of $t_{\mathrm{max}}$
was equal to the maximal time reached in time evolution ($t_{\mathrm{max}}=1200,1500,5000$). In this appendix, we discuss the impact of changes of $t_{\mathrm{min}}$ on the value of exponent $\overline \beta$. 

The result is shown in Fig.~\ref{figTmin}. For $W=8$, we observe that the value $\overline \beta$ remains, within the estimated error bars, constant in the interval $t_{\mathrm{min}}=[80,200]$, justifying the choice $t_{\mathrm{min}}=100$ made in the main text. Inclusion of times $t \lessapprox 80$ leads to an increase of $\overline \beta$ -- consistently with the small $t$ behavior of the imbalance shown in Fig.~\ref{figRDW8}. When  $t_{\mathrm{min}} \gtrapprox 200$, the precision of estimation of $\overline \beta$ decreases as the fitting interval $[t_{\mathrm{min}}, t_{\mathrm{max}}]$ gets narrower. Similar trends are observed for $W=10$ for data with  $t_{\mathrm{max}}=1500$. However, the stability of the fit is greatly improved when $t_{\mathrm{max}}=5000$: then, the choices of $t_{\mathrm{min}}$ from interval $[80,1000]$ lead to the values of $\overline \beta$
that agree within the estimated error bars.

The value of $\overline \beta$ decreases approximately $3$ times when $W$ is increased from $8$ to $10$. However,  $\overline \beta$ clearly remains positive within the estimated error bars, showing that the imbalance $I(t)$ indeed decays in time. Extrapolating the trend of changes in $\overline \beta$, we may expect that  $\overline \beta \approx 10^{-4}$ at $W=12$. Assuming a similar scaling of the statistical error of $\overline \beta$, already at $W=12$ we would need to either increase the number of disorder realizations or increase  $ t_{\mathrm{max}}$ as compared to their respective values at $W=8,10$ to be certain that the value of $\overline \beta$ at $W=12$ is positive.

  \begin{figure}[h]
  \centering
\includegraphics[width=0.91\linewidth]{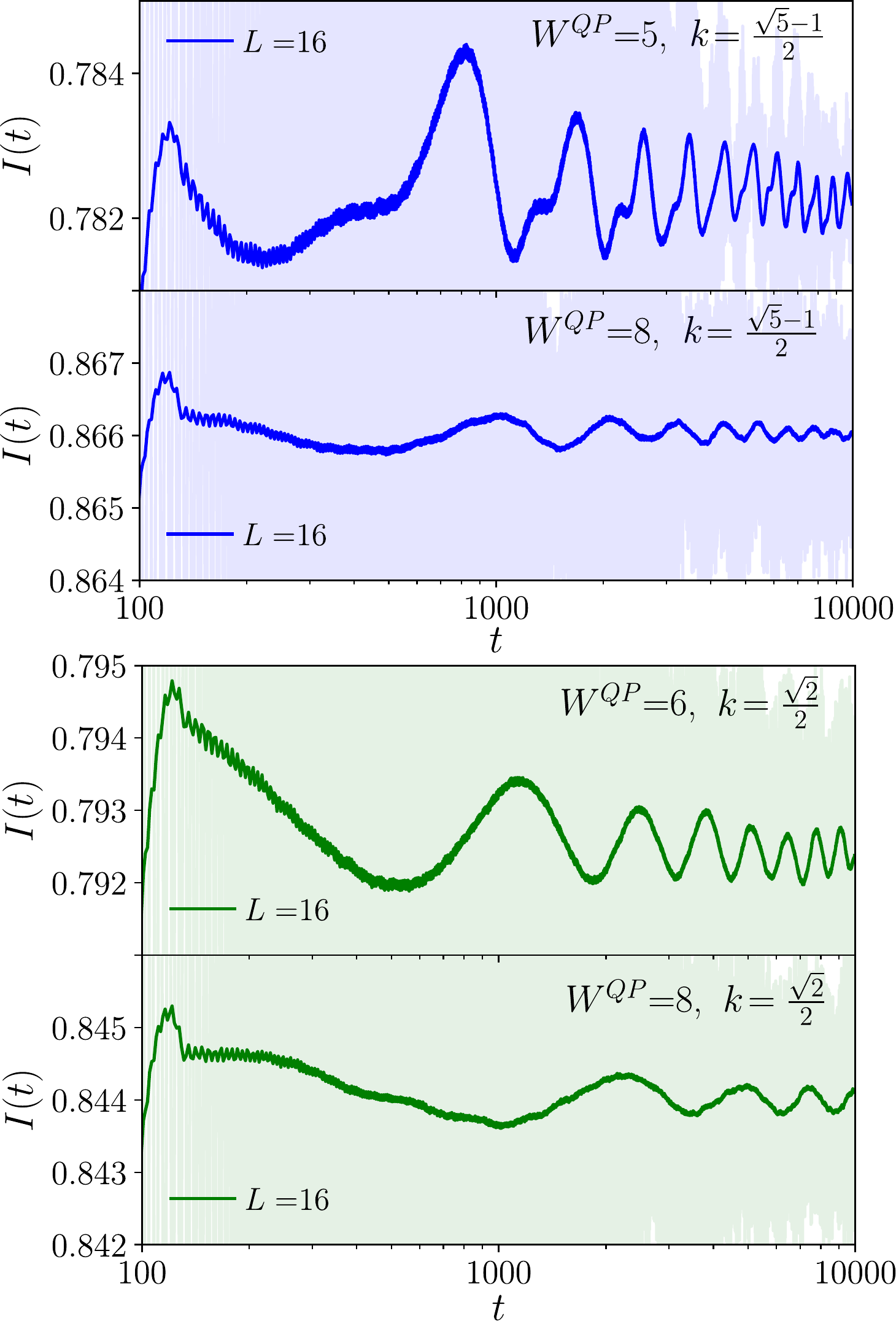}
\vspace{-0.4cm}
  \caption{{Persistent oscillations for QP potential. The imbalance $I(t)$ (averaged over times $[t-25,t+25]$) is shown by solid lines for various amplitudes of QP potential $W^{QP}$, shades show the imbalance without time averaging. The results are averaged over more than $5000$ realizations of QP potential and the system size is fixed as $L=16$. Top panel shows results for $k=
  \frac{\sqrt{5}-1}{2}$ whereas the bottom panel for $k=\frac{\sqrt{2}}{2}$. The range of the vertical axis is the same for all subplots.}
  }
  \label{figOsc}
\end{figure}  
\section{Oscillations of the imbalance for quasiperiodic systems  }
\label{app:osc}

{

In  Sec.~\ref{sec:imbaQP} of the main text, we have demonstrated {an} emergence of persistent oscillations of the imbalance $I(t)$ for sufficiently strong QP potential. In this Appendix we provide further details on this phenomenon. 

 Since the oscillations do not depend on the system size (at least for $L \geq 12$), we fix the system size as $L=16$ and investigate the time evolution of the imbalance $I(t)$ varying the amplitude $W^{QP}$ of the QP potential as well as the wave vector $k$ that determines the shape of the QP potential (recall that $h_j = W^{\mathrm{QP}} \cos(2\pi k j + \phi)$). The results are shown in Fig.~\ref{figOsc}. 
 
 By comparing the results for fixed $k=\frac{\sqrt{5}-1}{2}$, we note that the amplitude of oscillations  diminishes when $W^{QP}=5$ is increased to $W^{QP}=8$. This could be expected as in the limit of $W^{QP} \to \infty$, the initial N\'eel state becomes an eigenstate of the XXZ model. In that limit, the oscillations are absent and the imbalance remains trivially equal to unity throughout the time evolution. Thus, the imbalance oscillations occur only in a limited range of amplitudes of the QP potential: $W^{QP}$ 
 must be sufficiently large to give rise to a very slow dynamics (unlike in Fig.~\ref{figQPD1}) but not large enough to give rise to a trivial dynamics.
 A similar decrease of the oscillations of $I(t)$ upon the increase of $W^{QP}$ is visible in Fig.~\ref{figOsc} for $k=\frac{\sqrt{2}}{2}$. 
 
 The differences in the pattern of oscillations for $k=\frac{\sqrt{5}-1}{2}$ and $k=\frac{\sqrt{2}}{2}$ (well pronounced for the smaller values of $W^{QP}$ in Fig.~\ref{figOsc}) demonstrate that the oscillations of $I(t)$ depend in a non-trivial fashion on the value of the constant $k$. By investigating time dynamics of the density correlation function $C(t)$ for initial states that are eigenstates of the $S^z_i$ operator but are different than the N\'eel state, we noted that the pattern of the oscillations of $I(t)$ depends strongly also on the initial state. In particular, the density correlation function $C(t)$ averaged over such initial states shows no long-time oscillations. This shows that the emergence of the pattern of oscillations is determined by the interplay of a spatial structure of the initial state and the constant $k$ of the QP potential. 

}

%


\end{document}